\documentclass[nolinenumbers]{aastex631}

\usepackage{newtxtext,newtxmath}
\usepackage[T1]{fontenc}
\usepackage{natbib}
\usepackage{threeparttable}
\usepackage{booktabs,makecell,multirow}
\usepackage{longtable}
\usepackage{makecell}
\usepackage{diagbox}
\DeclareRobustCommand{\VAN}[3]{#2}
\let\VANthebibliography\thebibliography
\def\thebibliography{\DeclareRobustCommand{\VAN}[3]{##3}\VANthebibliography}
\usepackage{graphicx}
\graphicspath{{./}} % Including figure files
\usepackage{amsmath}	% Advanced maths commands

\newcommand
    {\comment}
    [1]
    {\textcolor
        {blue}
        {\textbf{#1}}
    }
    
\newcommand{\ps}[0]{\href{https://github.com/NickSwainston/pulsar_spectra}{\texttt{pulsar\_spectra}}}
\renewcommand{\textbf}[1]{#1}

\begin{document}

\title{Statistical analysis of pulsar flux density distribution}

\author[0009-0003-4524-6530]{H.W. Xu}
\affiliation{Guizhou Normal University, Guiyang 550001, China}

\author[0000-0002-1243-0476]{R.S. Zhao*}
\affiliation{Guizhou Normal University, Guiyang 550001, China}

\author{Erbil G\"{u}gercino\u{g}lu}
\affiliation{National Astronomical Observatories, Chinese Academy of Sciences, 20A Datun Road, Chaoyang District, Beĳing 100101, China}

\author{H. Liu}
\affiliation{Guizhou Normal University, Guiyang 550001, China}

\author{D. Li}
\affiliation{National Astronomical Observatories, Chinese Academy of Sciences, 20A Datun Road, Chaoyang District, Beĳing 100101, China}
\affiliation{School of Astronomy and Space Science, University of Chinese Academy of Sciences, Beĳing, 100049, China}
\affiliation{Computational Astronomy, Zhejiang Laboratory, Hangzhou 311121, China}
\affiliation{NAOC-UKZN Computational Astrophysics Centre, University of KwaZulu-Natal, Durban 4000, South Africa}

\author{P. Wang}
\affiliation{National Astronomical Observatories, Chinese Academy of Sciences, 20A Datun Road, Chaoyang District, Beĳing 100101, China}
\affiliation{Institute for Frontiers in Astronomy and Astrophysics, Beĳing Normal University, Beĳing 102206, China}

\author{C.H. Niu}
\affiliation{Central China Normal University, Wuhan 430079, China}

\author{C. Miao}
\affiliation{Zhejiang Lab, hangzhou 311121, China}

\author{X. Zhu}
\affiliation{Guizhou Normal University, Guiyang 550001, China}

\author{R.W. Tian}
\affiliation{Guizhou Normal University, Guiyang 550001, China}

\author{W.L. Li}
\affiliation{Guizhou Normal University, Guiyang 550001, China}

\author{S.D. Wang}
\affiliation{Guizhou Normal University, Guiyang 550001, China}

\author{Z.F. Tu}
\affiliation{Guizhou Normal University, Guiyang 550001, China}

\author{Q.J. Zhi}
\affiliation{Guizhou Normal University, Guiyang 550001, China}

\author{S.J. Dang}
\affiliation{Guizhou Normal University, Guiyang 550001, China}

\author{L.H. Shang}
\affiliation{Guizhou Normal University, Guiyang 550001, China}

\author{S. Xiao}
\affiliation{Guizhou Normal University, Guiyang 550001, China}

\correspondingauthor{R.S. Zhao}
\email{201907007@gznu.edu.cn}

\begin{abstract}

This study presents a comprehensive analysis of the spectral properties of 886 pulsars across a wide frequency range from 20MHz to 343.5GHz, including a total of 86 millisecond pulsars. The majority of the pulsars exhibit power-law behavior in their spectra, although some exceptions are observed. Five different spectral models, namely simple power-law, broken power-law, low-frequency turn-over, high-frequency cut-off, and double turn-over, were employed to explore the spectral behaviors. The average spectral index for pulsars modeled with a simple power-law is found to be $-1.64 \pm 0.80$, consistent with previous studies. Additionally, significant correlations between the spectral index and characteristic parameters are observed particularly in millisecond pulsars, while no strong correlation is observed in normal pulsars. Different models show variations in the most influential characteristic parameters associated with the spectral index, indicating diverse dominant radiation mechanisms in millisecond pulsars.Finally, this study identifies 22 pulsars of the Gigahertz-peaked Spectra (GPS) type for the first time based on the Akaike information criterion.

\end{abstract}


%%%%%%%%%%%%%%%%% BODY OF PAPER %%%%%%%%%%%%%%%%%%

\section{Introduction}

Pulsars have been a topic of great interest to researchers since their discovery in 1968 by \citet{Hewish1968} because of their unique characteristics. With the development of observational techniques and tools, to date more than 3400 pulsars have been discovered\footnote{\url{https://www.atnf.csiro.au/research/pulsar/psrcat/}}. Previous studies suggested that the spectra of pulsars can be fitted with a simple power law, with normal pulsars having a flatter spectrum and millisecond pulsars having a steeper spectrum \citep{kramer1998}. As for the power-law of pulsar spectrum, the detection of pulsar intensity is more difficult with increasing frequency. Furthermore, low-frequency observations often fall within the range of a few hundred MHz (\citealp[a]{Manchester1978a}; \citealp{Camilo1995, Kondratiev2016, McEwen2020, Lee2022}), while high-frequency observations above $8$ GHz are relatively scarce ( \citet{Morris1981,Maron2004,Maron2013,Johnston2006,Zhao2019} and \citet[b]{Kramer1997b}).

\citet{Sieber1973} conducted extensive study on spectral index by collecting flux densities of $27$ pulsars and obtained their spectral indices in the range between $-0.7$ and $-3.3$. He found that the spectral behavior of pulsars exhibits distinct patterns and identified four different spectral behavior model: simple power-law model, broken power-law model, high-frequency cut-off model, and low-frequency turn-over model. Many studies have consistently favored the use of a simple power-law model to describe the relationship between the flux density and the observed frequency for most pulsars (\citealp{Sieber1973,Morris1981,Maron2013, McEwen2020, Lee2022}). Later, \citet{Malofeev1994} explored the relationship between spectral indices before and after inversion in the broken power-law model. Good agreement was found when comparing the predictions of the coherent curvature radiation model with their spectral data. Despite this, there is currently no clear model that can fully explain the diversity of spectral behavior.
Recently,  \citet{Spiewak2022} carried out a study on the spectral indices of $189$ pulsars. The results indicated an average spectral index of $-1.92$. Additionally, \citet{Lee2022} discussed the spectral indices of $22$ pulsars using SKA for the first time. This study revealed that pulsars with precisely determined spectral indices were more prone to demonstrated spectral turn-over phenomena. This highlights the importance of studying spectral indices in gaining a better understanding of emission mechanisms of pulsars.

The relationship between the spectral index and characteristic parameters of pulsars has been widely discussed since \citet{lorimer1995}. They found a correlation between the spectral index of pulsars and their characteristic age: young pulsars were observed to have predominantly flatter spectra compared to older pulsars. Subsequently, \citet{kramer1998} directed their focus to millisecond pulsars, a distinct group of pulsars. The spectral index of millisecond pulsars is steeper than that of normal pulsars, as previously reported by \citet{toscano1998}. In recent years, \citet{Han2016,jankowski2018,Zhao2019} have also studied the correlation between different characteristic parameters of pulsars, which will be further discussed in Section \ref{sec:discussion}.

Additionally, the research on the spectral index has led to the discovery of a new class of pulsars called GPS. As noted by \citet{kijak2011b} and \citet{dembska2014}, GPS pulsars, unlike other pulsars, exhibit flipping phenomena and peak flux density that occur approximately at $1$ GHz. Therefore, studying the spectral indices of GPS pulsars is of significant interest.

Nevertheless, inaccurate spectral index values for many pulsars still exist, which is mainly attributed to the lack of high-precision spectral data. Therefore, studying pulsar spectral indices necessitates an extensive literature review to extract the essential data, which is a time-consuming and error-prone process. Furthermore, only a small number of empirical models are currently available because no complete theoretical model exists to accurately fit the wide spectrum of spectral indices found in pulsars.

The purpose of this paper is to reanalyze the spectral information of $886$ pulsars using published literature and publicly available data on the ATNF. \textbf{The paper is organized as follows. In Section \ref{sec:data processing}, we outline the methodology used for data processing. In Section \ref{sec:results}, we present the statistical results and correlation analysis. The discussion and conclusions are given in Section \ref{sec:discussion} and Section \ref{sec:conclusions}, respectively.}

\section{Data processing}
\label{sec:data processing}

\textbf{Based on the original dataset provided by \citet{swainston2022} which included 34 publications in their catalogue (see Table 1 in that paper), we meticulously collected and compiled all data on pulsar flux density from publicly available online sources and published literature. The resulting extended data on pulsar spectra, which enlarged the original dataset by 50 percent, are referenced in Table \ref{tab:tablefre0}. Both the original data \citep{swainston2022} and the extended data (this work) must satisfy some specific requirements. For instance, any paper that includes the data must undergo a rigorous process of peer review and be accepted for publication. Additionally, it is necessary for the data to encompass information concerning frequency bandwidth and other related factors\footnote{\url{https://pulsar-spectra.readthedocs.io/en/latest/catalogue.html}}. We then proceeded to contribute this extended dataset to the \ps{}\footnote{\label{first-link}\url{https://github.com/NickSwainston/pulsar_spectra}} database, hosted on GitHub and designed by \citet{swainston2022}.
This is an open-source catalog of pulsar flux density and automated spectral fitting software that can find the best spectral models. It is based on the Python language, allowing users to add new spectral measurement results to the catalog upon publication (thus providing access to the latest version through the \emph{\comment{homepage}} \textsuperscript{\ref{first-link}}). Here, we will only provide a brief overview of the \ps{} database, and more detailed information can be found on their \emph{\comment{GitHub page}} \textsuperscript{\ref{first-link}}.}

\setcounter{table}{0}

\begin{longtable}{lcc}
\caption{\textbf{Papers included in our catalogue.}}  
\label{tab:tablefre0}\\
\toprule
Paper& Pulsars& Freq. range (MHz) \\
\midrule
\endfirsthead
\multicolumn{3}{c}{{\tablename} \thetable{} (Continued)} \\
\toprule
Paper& Pulsars& Freq. range (MHz) \\
\midrule
\endhead
\bottomrule
\multicolumn{3}{r}{{Continued on the next page}} \\
\endfoot
\bottomrule
\endlastfoot
\citet{aloisi2019green}       & 4          & 350-350     \\
\citet{bailes1997discovery}        & 4          & 400-1400    \\
\citet{basu2018gigahertz}         & 6          & 325-1280    \\
\citet{bhat2023southern}           & 120        & 154-154     \\
\citet{biggs1996search}         & 4          & 408-408     \\
\citet{boyles2013green}         & 13         & 820-820     \\
\citet{brinkman2018no}       & 12         & 327-1400    \\
\citet{champion2005a}      & 17         & 430-430     \\
\citet{champion2005b}      & 1          & 327-430     \\
\citet{crawford2001polarization}        & 9          & 660-2264    \\
\citet{crawford2007flux}        & 2          & 1384-3100   \\
\citet{deller2009precision}          & 9          & 1650-1650   \\
\citet{dembska2015flux}         & 6          & 610-610     \\
\citet{demorest2012limits}      & 17         & 327-2300    \\
\citet{esamdin2004study}          & 2          & 327-327     \\
\citet{freire2007timing}          & 1          & 350-1950    \\
\citet{gentile2018nanograv}         & 28         & 430-2100    \\
\citet{giacani2001pulsar}          & 2          & 1420-8460   \\
\citet{han1999pulsars}             & 106        & 1435-1435   \\
\citet{von1997effelsberg}     & 27         & 1410-10450  \\
\citet{joshi2009discovery}          & 3          & 626-1400    \\
\citet{kaspi1997discovery}       & 1          & 660-1650    \\
\citet{kijak1998pulse}         & 87         & 4850-4850   \\
\citet{kramer1997observations}       & 4          & 14600-43000 \\
\citet{kuniyoshi2015low}       & 10         & 74-1400     \\
\citet{lewandowski2004}     & 18         & 430-1400    \\
\citet{lorimer1995}        & 4          & 436-1520    \\
\citet{lorimer1996}         & 4          & 436-1400    \\
\citet{lorimer2005}        & 38         & 400-430     \\
\citet{lorimer2007}         & 1          & 430-430     \\
\citet{lynch2012}           & 12         & 2000-2000   \\
\citet{lynch2013}          & 10         & 820-820     \\
\citet{manchester1995}      & 2          & 1400-8300   \\
\citet{manchester2013}     & 20         & 700-3100    \\
\citet{michilli2020}        & 19         & 129-1532    \\
\citet{mickaliger2012}       & 1          & 820-820     \\
\citet{mikhailov2016}      & 2          & 146-146     \\
\citet{ng2015}            & 57         & 325-1352    \\
\citet{rozko2018}          & 2          & 325-5900    \\
\citet{sayer1997}           & 8          & 370-800     \\
\citet{seiradakis1995}    & 188        & 1315-10550  \\
\citet{shapiro2021} & 3          & 430-1400    \\
\citet{stovall2014}         & 67         & 350-820     \\
\citet{surnis2018}         & 3          & 325-1170    \\
\citet{titus2019}          & 3          & 1382-1382   \\
\citet{Zhao2017}      & 26         & 8600-8600   \\

\end{longtable}
\begin{threeparttable}
\end{threeparttable}

\subsection{Spectral index correlations}

\textbf{Five models, namely the simple-power law, broken power law, low-frequency turn-over, high-frequency cut-off, and double turn-over spectrum have been used in \ps{}, to describe the spectral behavior with frequency. To determine the most suitable model among the five options, the software employs the methodology outlined by \citet{jankowski2018}. This approach relies on robust statistical techniques to determine the optimal fitting model for individual pulsar spectra, as detailed in \citet{swainston2022}. The five models are as follows (with $S_{\rm \nu}$ is the flux density at frequency $\nu$, $\nu_{\rm 0}$ is reference frequency, and $c$ is a constant):}

(i) Simple power law: 
\begin{equation}
    S_{\rm \nu} = c \left( \frac{\nu}{\nu_{\rm 0}} \right)^\alpha,
\end{equation}
where $\alpha$ is the spectral index.

\begin{figure}
	\includegraphics[width=\columnwidth]{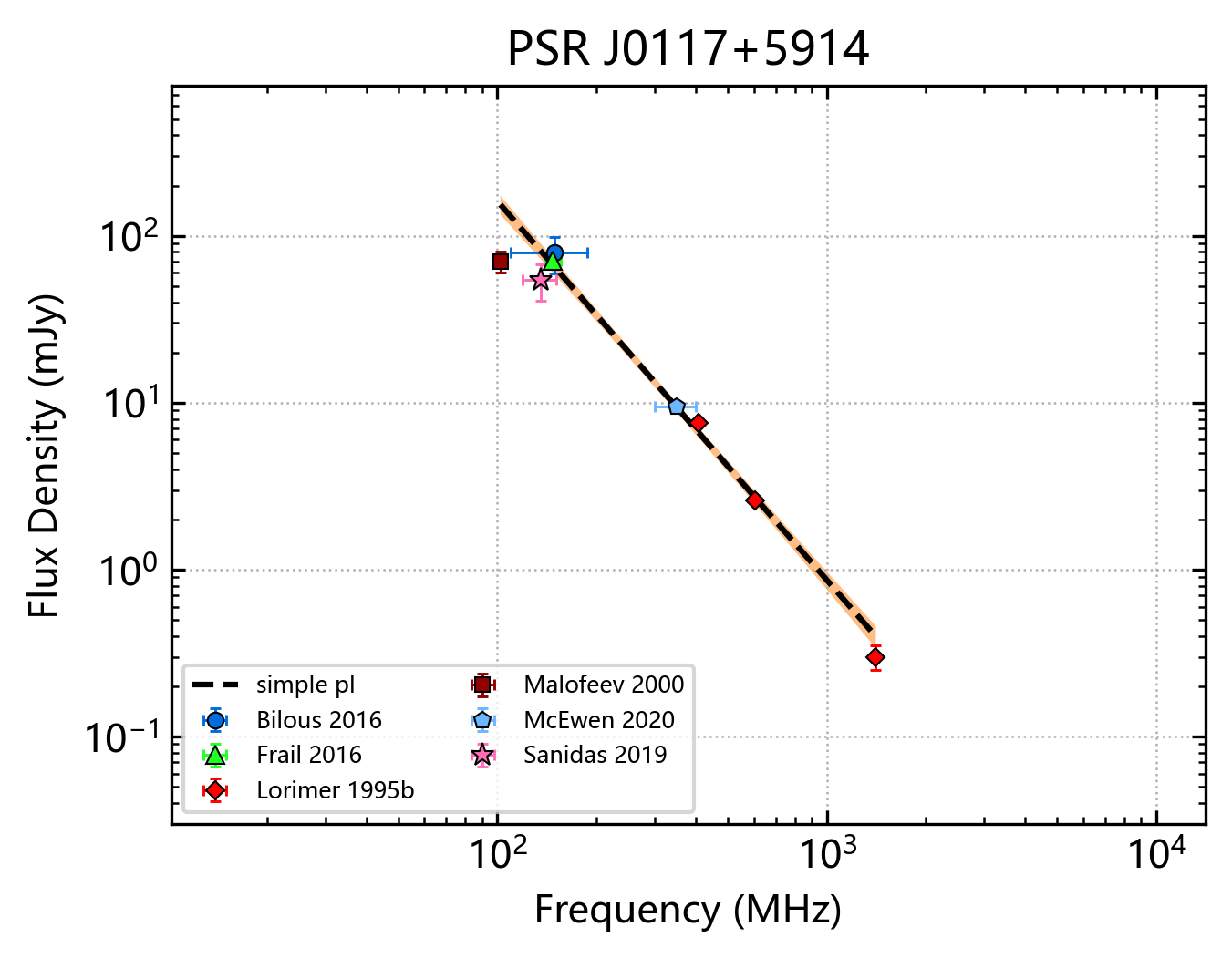}
    \caption{Simple power law model; Black dashed line: the best-
fitting model to the data. Orange shaded envelope: the 1$\sigma$ uncertainty of the best-fitting model.}
    \label{fig:J0117+5914}
\end{figure}

(ii)Broken power law:
\begin{equation}
    S_{\rm \nu} = c\begin{cases}
            \left( \frac{\nu}{\nu_{\rm 0}} \right)^{\alpha_1}   & \mathrm{if}\: \nu \leq \nu_{\rm b} \\[5pt]
            \left( \frac{\nu}{\nu_{\rm 0}} \right)^{\alpha_2} \left( \frac{\nu_{\rm b}}{\nu_{\rm 0}} \right)^{\alpha_1-\alpha_2} & \mathrm{otherwise} \\
        \end{cases},
\end{equation}
where $\nu_{\rm b}$ is the frequency of the spectral break, $\alpha_1$ and $\alpha_2$ are the spectral index before and after the break, respectively.

\begin{figure}
	\includegraphics[width=\columnwidth]{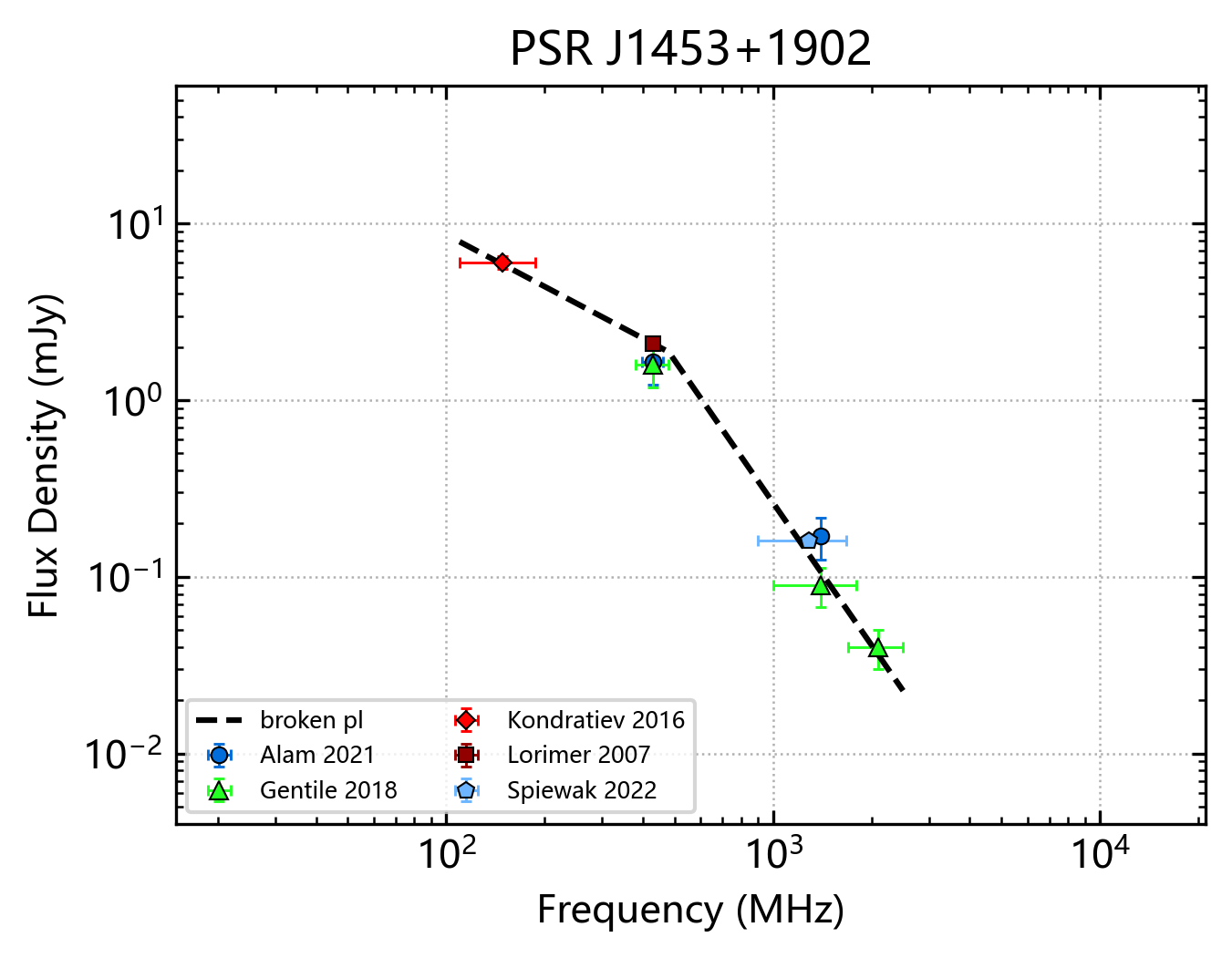}
    \caption{Broken power law model; Black dashed line: the best-
fitting model to the data.}
    \label{fig:J0040+5716}
\end{figure}

(iii)Power law with low-frequency turn-over:
\begin{equation}
    S_{\rm \nu} = c\left( \frac{\nu}{\nu_{\rm 0}} \right)^{\alpha} \exp\left [ \frac{\alpha}{\beta} \left( \frac{\nu}{\nu_{\rm peak}} \right)^{-\beta} \right ],
\end{equation}
where $\alpha$ is the spectral index, $\nu_{\rm peak}$is the turn-over frequency, and $0 < \beta \leq 2.1$ determines the smoothness of the turn-over.
\begin{figure}
	\includegraphics[width=\columnwidth]{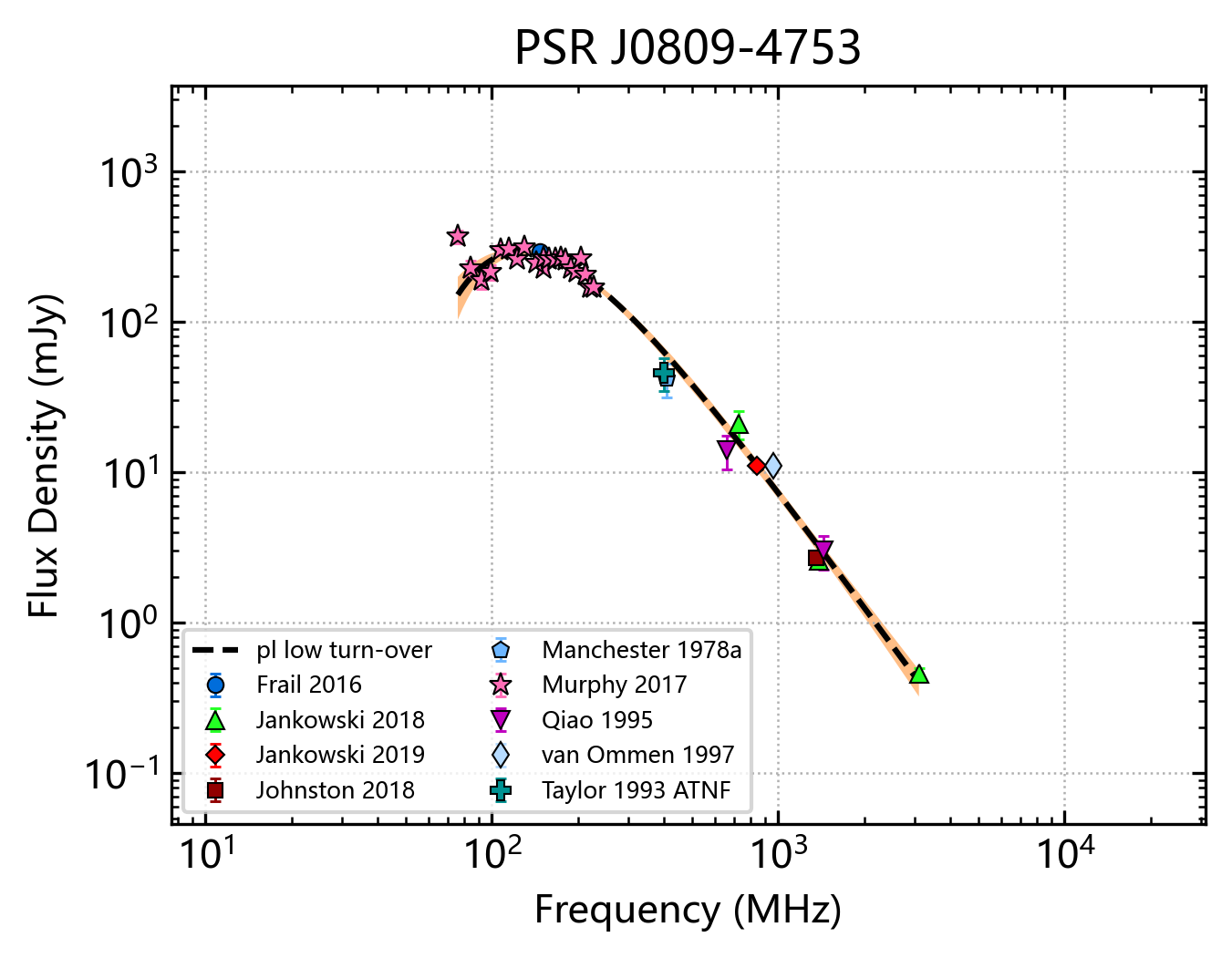}
    \caption{Low-frequency turn-over law model; Black dashed line: the best-
fitting model to the data. Orange shaded envelope: the 1$\sigma$ uncertainty of the best-fitting model.}
    \label{fig:J0809-4753}
\end{figure}

(iv)Power law with high-frequency cut-off:
\begin{equation}
    S_{\rm \nu} = c\left( \frac{\nu}{\nu_{\rm 0}} \right)^{\alpha} \left ( 1 - \frac{\nu}{\nu_{\rm c}} \right ),\qquad \nu < \nu_{\rm c},
\end{equation}
where $\alpha$ is the spectral index and $\nu_{\rm c}$ is the cut-off frequency.

\begin{figure}
	\includegraphics[width=\columnwidth]{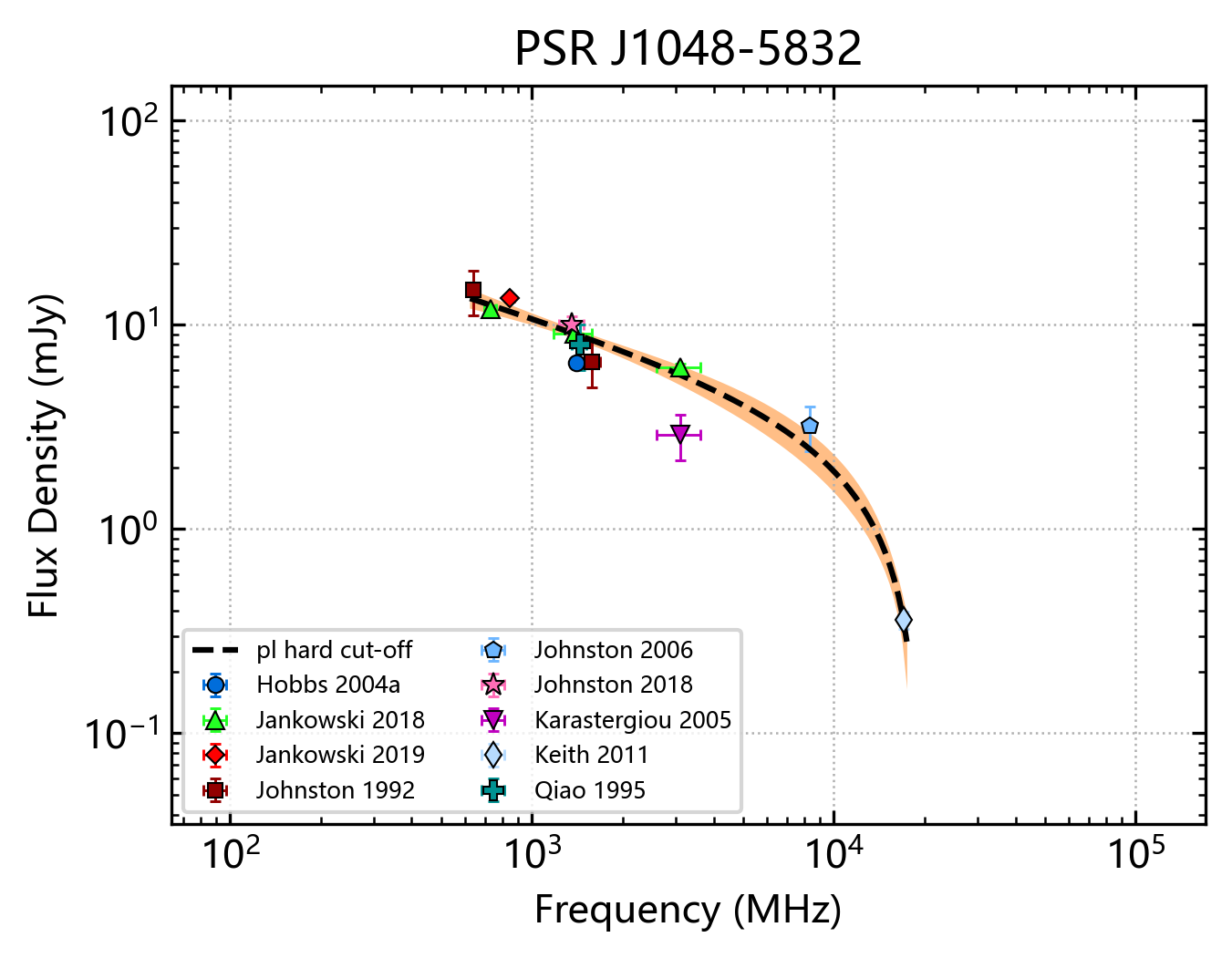}
    \caption{High-frequency cut-off model; Black dashed line: the best-
fitting model to the data. Orange shaded envelope: the 1$\sigma$ uncertainty of the best-fitting model.}
    \label{fig:J1048-5832}
\end{figure}

(v)Power law with double-turn-over:
\begin{equation}
    S_{\rm \nu} = c \left( \frac{v}{\nu_{\rm 0}} \right)^a \left ( 1 - \frac{v}{\nu_{\rm c}} \right ) \exp \left [ \frac{a}{\beta} \left( \frac{v}{\nu_{\rm c}} \right)^{-\beta} \right ], \quad \nu < \nu_{\rm c},
\end{equation}
where $\alpha$ is the spectral index, $\nu_{\rm peak}$ and $\nu_{\rm c}$ are the turn-over frequency and the cut-off frequency, respectively, and $0 < \beta \leq 2.1$ determines the smoothness of the turn-over.

\begin{figure}
	\includegraphics[width=\columnwidth]{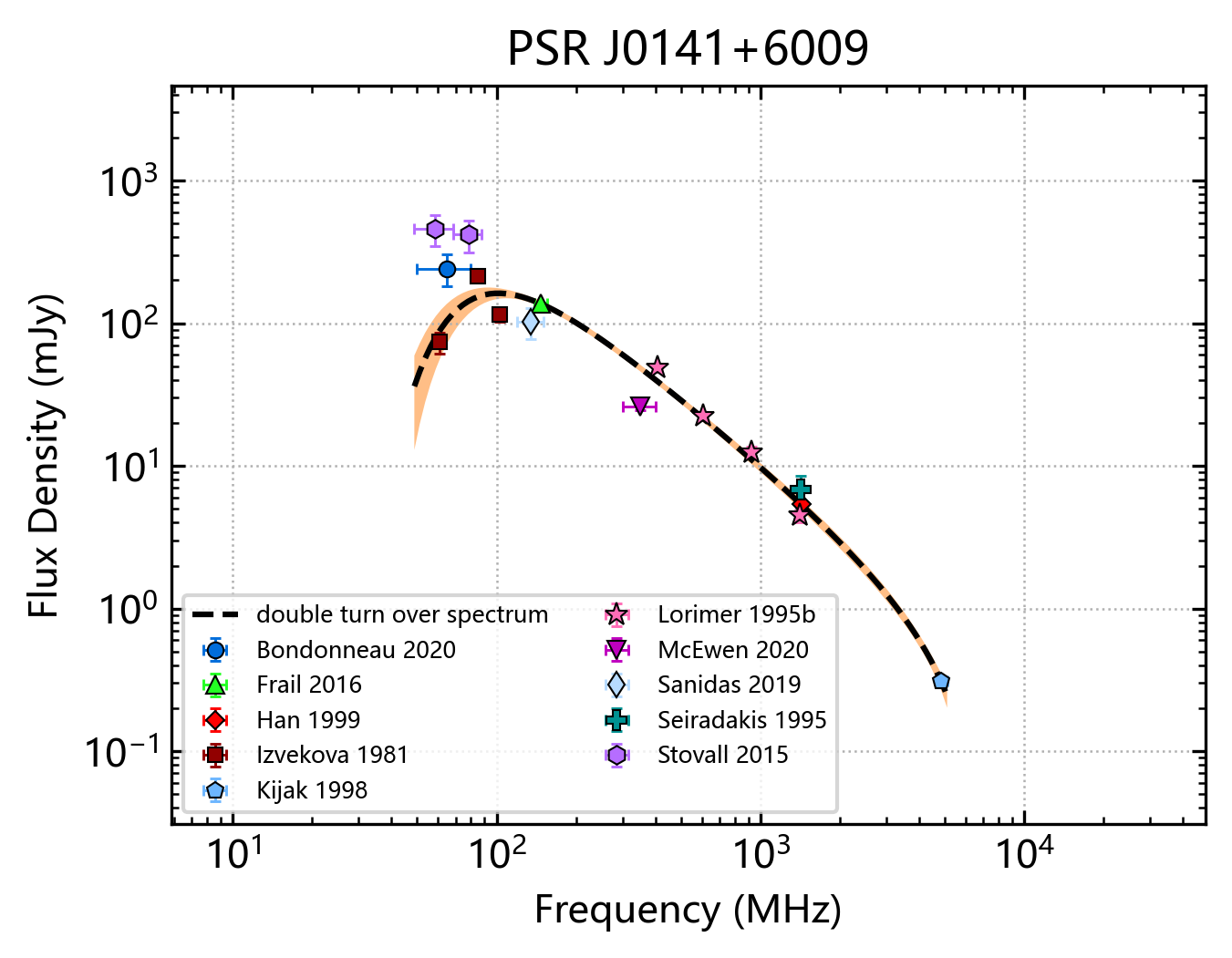}
    \caption{Double-turn-over law mode; Black dashed line: the best-
fitting model to the data. Orange shaded envelope: the 1$\sigma$ uncertainty of the best-fitting model.}
    \label{fig:J0141+6009}
\end{figure}

\textbf{Since the extended data compiled from different studies were obtained by various telescopes with different calibration programs and observation frequencies, errors are unavailable for certain flux density values, resulting in visible deviations in the modeling of certain pulsars.} Despite this, we treated such unmodelable pulsars as integral to the database's completeness, and we anticipate an improvement in the process through the collection of pertinent data or the updating of software.

To enhance the accuracy of the spectral index of pulsars, it is necessary to ensure that each pulsar includes flux density values at the four distinct frequencies. Accordingly, we chose 941 pulsar samples out of a selection of more than 3300 pulsars that satisfy this requirement. Next, \ps{} was employed to perform data modeling. \ps{} applies the five models to each pulsar, subsequently determining the final model. However, some pulsars cannot be fitted well by \ps{}. We separately fitted this portion of the data using a simple power law and provided an approximate spectral index. The fitting results for this portion are shown in \ref{tab:a5}. As a result, we performed a comprehensive analysis of the spectral information including 886 pulsars.

\section{Results and analysis}
\label{sec:results}

\setcounter{table}{1}
\begin{table*}
\caption{\textbf{The distribution of different spectral models in normal pulsars and millisecond pulsars. The first column represents the five models we have used. The second and fourth columns indicate the number of normal pulsars and millisecond pulsars in each model respectively. The following third and fifth columns represent the percentage of the pulsars corresponding to each model.}} 
\setlength{\tabcolsep}{8mm}
\centering 
\begin{threeparttable}
\label{tab:table00}
\begin{tabular}{lcccc} 
\toprule 
Set & Normal pulsars & percentage & MSP & percentage\\
\midrule 
Simple power-law model   & 553 & 69.13                 & 54  & 62.79                 \\
Broken power law model  & 45  & 5.63                 & 9   & 10.47             \\
High-frequency cut-off models  & 100 & 12.50                   & 7   & 8.14                  \\
Low-frequency turn-over models  & 86  & 10.75                 & 9   & 10.47                \\
Double turn-over model   & 16  & 2.00                   & 7   & 8.14                \\
Total &800 &-  & 86&- \\
\bottomrule 
\end{tabular}
\end{threeparttable}
\end{table*}

Among the identified pulsars, around 90\% of sources correspond to normal pulsars, amounting to a total number of 800. The majority, at 69.13\%, obey the simple power-law model, while 12.50\% and 10.75\% conform to the high-frequency cut-off and low-frequency turn-over models, respectively. The least eligible models are the broken power-law model and double turn-over model with amounting to merely 5.63\% and 2.00\% of the total, respectively. Millisecond pulsars constitute roughly 10\% of the entire population with 86 identified instances. Within this subset, the simple power-law model retains the highest correlation coefficient at 62.79\%, followed by the broken power-law and low-frequency turn-over models, both being 10.47\%, and finally, the high-frequency cut-off and double turn-over models at 8.14\%(see Table \ref{tab:table00}). We presented our investigation of the non-simple power law models in tables, with \ref{tab:a1} offering insights on the broken power law model and \ref{tab:a3} showing the low-frequency turn-over model results. \ref{tab:a4} and \ref{tab:a2} detailed the findings of high-frequency cut-off and the double turn-over models, respectively. We also mark the MSPs with asterisk in the table. Particularly, we present the spectral analysis of GPS pulsars in Section \ref{sec:gps}. According to \cite{kijak2011b}, GPS pulsars are believed to be relatively young and have high dispersion measures (DM), which may be related to their surrounding environment.

\subsection{Normal pulsars}

\begin{figure}
	\includegraphics[width=\columnwidth]{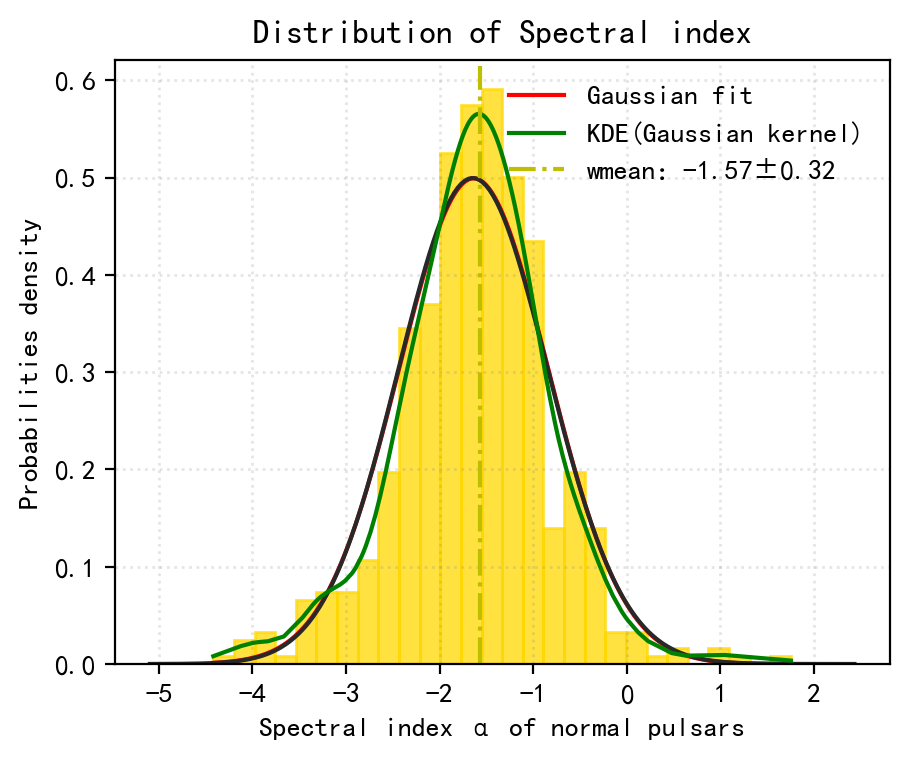}
    \caption{Histogram of the spectral indices $\alpha$ for normal pulsars that were classified to have simple power-law. We display a Gaussian and a kernel density estimation (KDE) using a Gaussian kernel fit to the data with a red and green line. We show a weighted mean spectral index\textbf{(wmean: $-1.57 \pm 0.32$)}. }
    \label{fig:3.1.1.1}
\end{figure}

\subsubsection{Simple power-law model}

we have identified a total of $553$ normal pulsars which can be effectively modeled using a simple power-law. The corresponding spectral histogram of the resulting analysis, is shown in Figure \ref{fig:3.1.1.1}.

 We calculated the average spectral index of $-1.57 \pm 0.32$ by weighting their reciprocal errors for each pulsar, and estimated the uncertainty by standard deviation. The weighted mean spectral index from \citet{jankowski2018} is consistent with our results within uncertainties. In addition, we constructed a kernel density estimation (KDE) plot(green line) and a Gaussian fitting(red line)  \textbf{in Figure \ref{fig:3.1.1.1}}. \textbf{It is worth noting that eleven normal pulsars show positive spectral indices, excluding the pulsar PSR J1745-2900 for which high-frequency data is lacking.} Furthermore, we performed the KS test and obtained a p-value of $0.18$.

\subsubsection{Broken power-law spectra model}
\label{subsec:bplnp}

\begin{figure}
	\includegraphics[width=\columnwidth]{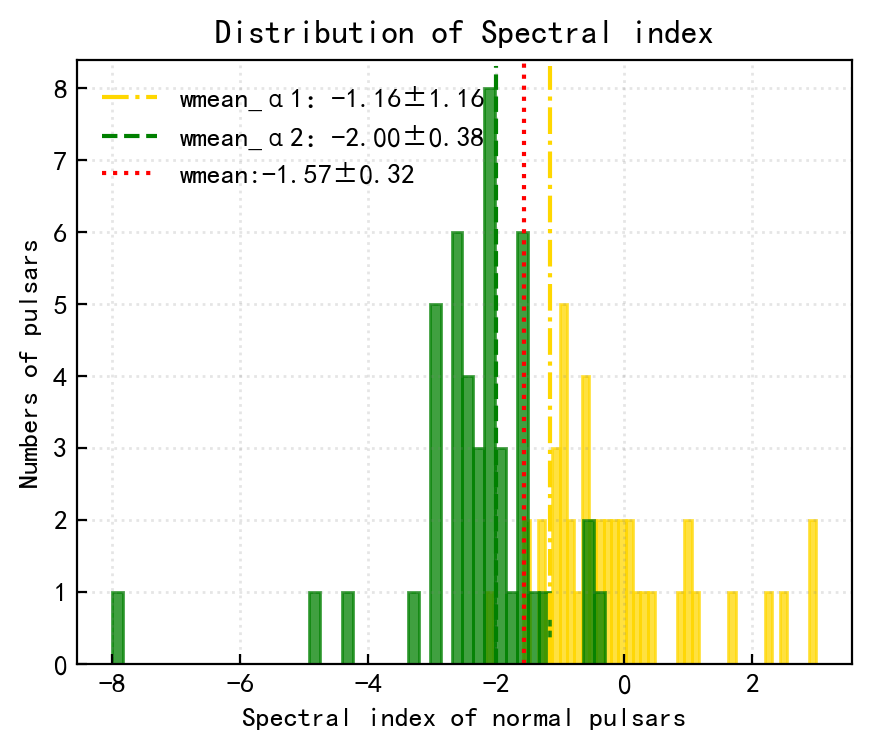}
    \caption{Histogram of the spectral indices $\alpha$ for normal pulsars that were classified to have broken power-law. Among them, green and yellow represent the spectral indices before and after the break, respectively. The corresponding dashed lines of the colors represent the corresponding weighted averages. In addition, we specifically use a red dashed line to represent the weighted average of normal pulsars in the simple power-law model.}
    \label{fig:3.1.2.1}
\end{figure}

The broken power-law model was a rare occurrence, with only $5.63\%$ of the normal pulsars in Figure \ref{fig:3.1.2.1} exhibits such a behaviour. Based on the fitting results, histograms of the spectral indices were plotted. The weighted average spectral indices before and after the broken are $-1.16 \pm 1.16$ and $-2.00 \pm 0.38$, respectively.

\begin{figure}
	\includegraphics[width=\columnwidth]{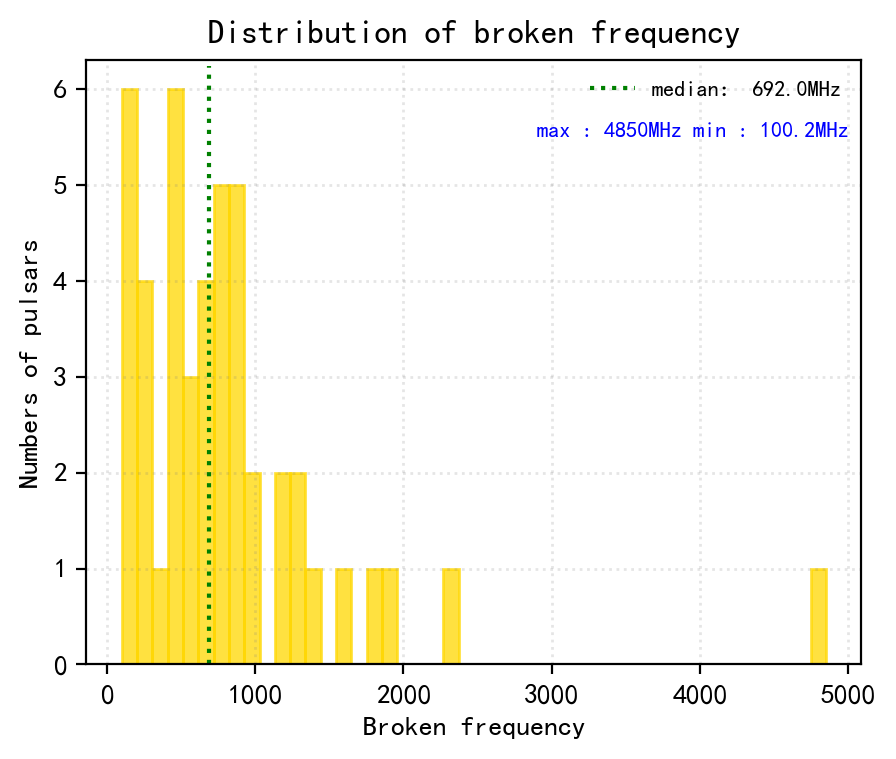}
    \caption{The frequency histogram at the points of spectral breaks. We have marked the position of the median with a dashed line in the figure, and provided its value, as well as the maximum and minimum values, in the upper right corner.}
    \label{fig:3.1.2.2}
\end{figure}

Furthermore, to discuss the location of spectral breaks, we constructed a frequency histogram at the points of spectral breaks in Figure \ref{fig:3.1.2.2}. The results reveal that all pulsars, except for J1705-1906 at $4850$ MHz and J1935+1616 at $2291$ MHz, exhibit spectral breaks below $2$ GHz. Upon examination of the break frequencies, the median value (dashed line) is observed $692$ MHz. The maximum position of the spectral break is $4850$ MHz, and the minimum is  $100$ MHz. Consequently, it can be seen from the figure that the breaks are mainly concentrated within the $100$ MHz to $1$ GHz range.

\begin{figure}
	\includegraphics[width=\columnwidth]{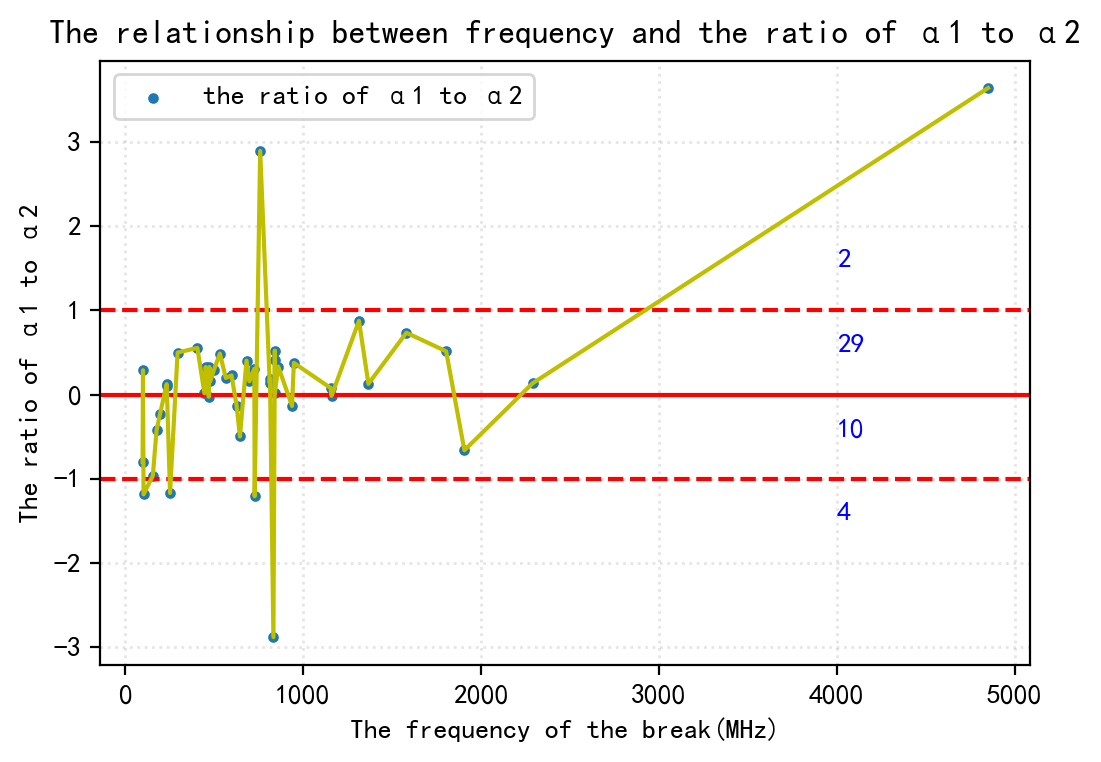}
    \caption{The relationship between frequency and the ratio of \textbf{$\alpha_1$ to $\alpha_2$, where $\alpha_1$ and $\alpha_2$ are the spectral indices before and after the break}. The blue numbers indicate the number of pulsars in each region.}
    \label{fig:3.1.2.3}
\end{figure}

Figure \ref{fig:3.1.2.3} shows a scatter plot of the ratio between the spectra before and after the spectral break against the position of the break. Since some pulsars exhibit positive spectral indices before the break but negative spectral indices after the break, the graph can be divided into four regions. These regions, delimited by solid and dashed red lines, are determined by the spectral index ratio. Within the range of -1 to 1, a total of 33 pulsars have spectral indices that get flatten after the break, while the remaining pulsars exhibit steeper spectral indices. Notably, among the 14 pulsars with negative values, their spectral indices transition from positive to complex values, indicating a flipping behavior in their spectral behavior. Therefore, the broken power-law model reveals three distinct spectral behaviour patterns. Among these patterns, 29 pulsars exhibit flatter spectral indices after the break, while 2 pulsars exhibit steeper spectral indices. The remaining 14 pulsars display a flipping behavior in their spectral indices.

\subsubsection{Power-law spectra with low-frequency turn-over model}

The PULSAR-SPECTRA model employs a synchrotron radiation function to describe the low-frequency turn-over process \citep{izvekova1981}, but can describe the spectra expected due to both synchrotron self and thermal free–free absorption. In this model, a variable $\beta$ is utilized as an unspecified parameter, which determines the smoothness of the turn-over; when it equals to $2.1$, 
 $\beta$ can represent free-free absorption, as indicated by \cite{rajwade2016} and \cite{kijak2017}. Therefore, the outcomes of the low-frequency turn-over model indicate that approximately $41.8\%$ of absorption results from a free-free absorption model. we specifically used the Kolmogorov-Smirnov test to analyze the distribution of spectral indices in these two scenarios. The results indicate that they originate from different distributions.

\begin{figure}
	\includegraphics[width=\columnwidth]{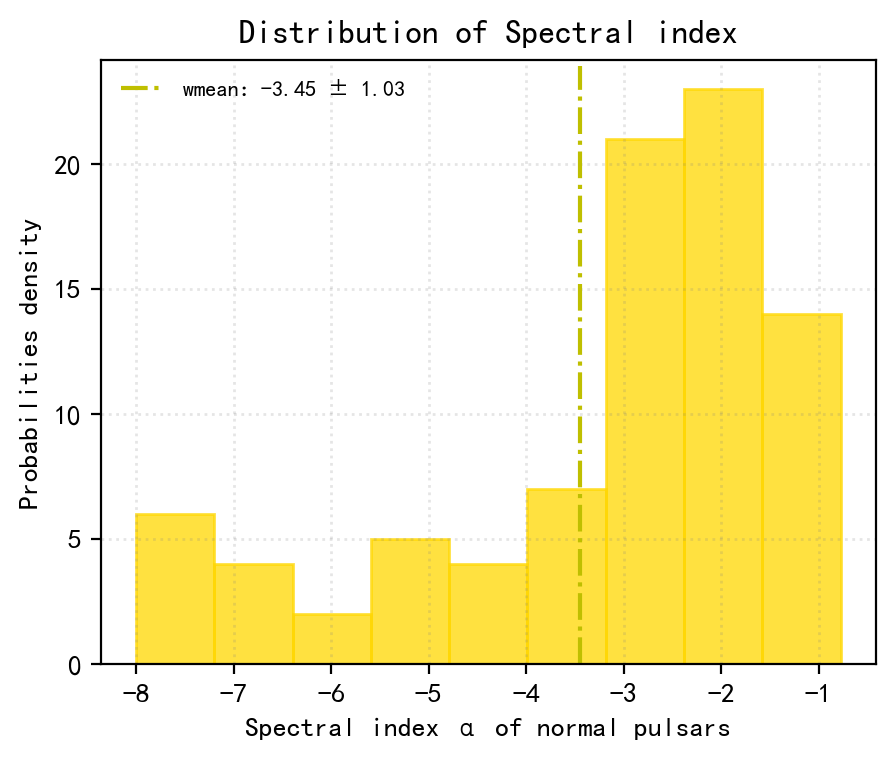}
    \caption{Histogram of the spectral indices $\alpha$ for normal pulsars that were classified to have low-frequency turn-over. The dashed line represents the weighted average.}
    \label{fig:3.1.3.1}
\end{figure}

\begin{figure}
	\includegraphics[width=\columnwidth]{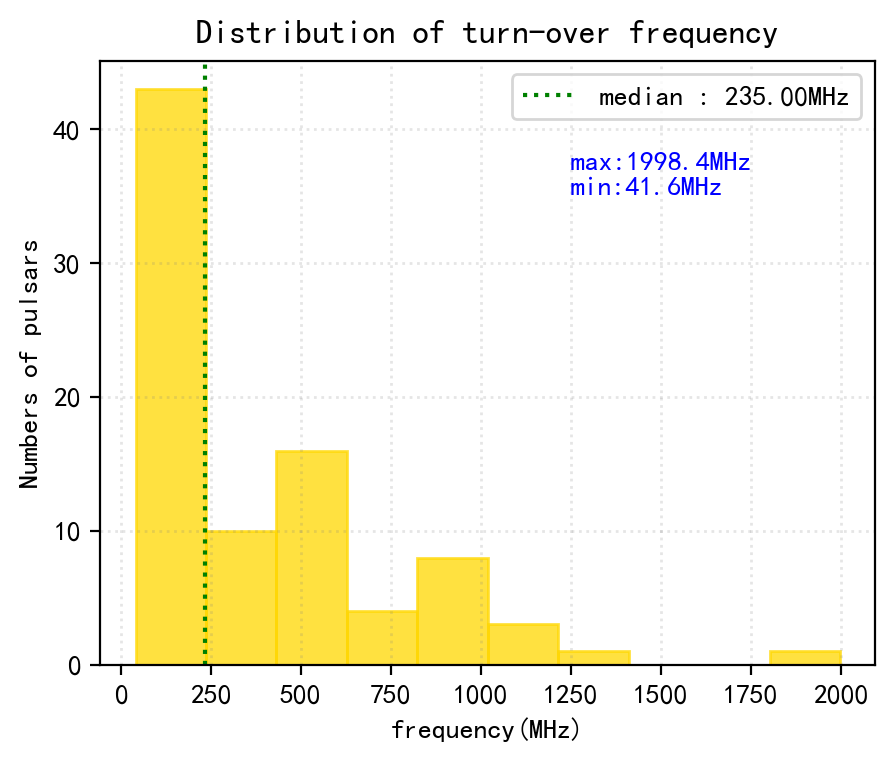}
    \caption{Distribution of turn-over frequency. We also have marked the position of the median with a dashed line in the figure, and provided its value, as well as the maximum and minimum values, in the upper right corner.}
    \label{fig:3.1.3.2}
\end{figure}

In terms of the distribution of spectral indices (as shown in Figure \ref{fig:3.1.3.1}), the low-frequency turn-over has a weighted average of $-3.59\pm1.13$ indicating a smaller value in contrast to typical spectral index values. Additionally, concerning the turn-over frequency distribution (shown in the Figure \ref{fig:3.1.3.2}), the maximum frequency recorded was $1998.4$ MHz, and the minimum frequency was $41.6$ MHz. The median frequency recorded was $235$ MHz (dashed line). This value is closer to the minimum value rather than the maximum value, indicating that the locations of low-frequency turn-over are more concentrated below 235 MHz.

\subsubsection{Power-law spectra with high-frequency cut-off model}

\begin{figure}
	\includegraphics[width=\columnwidth]{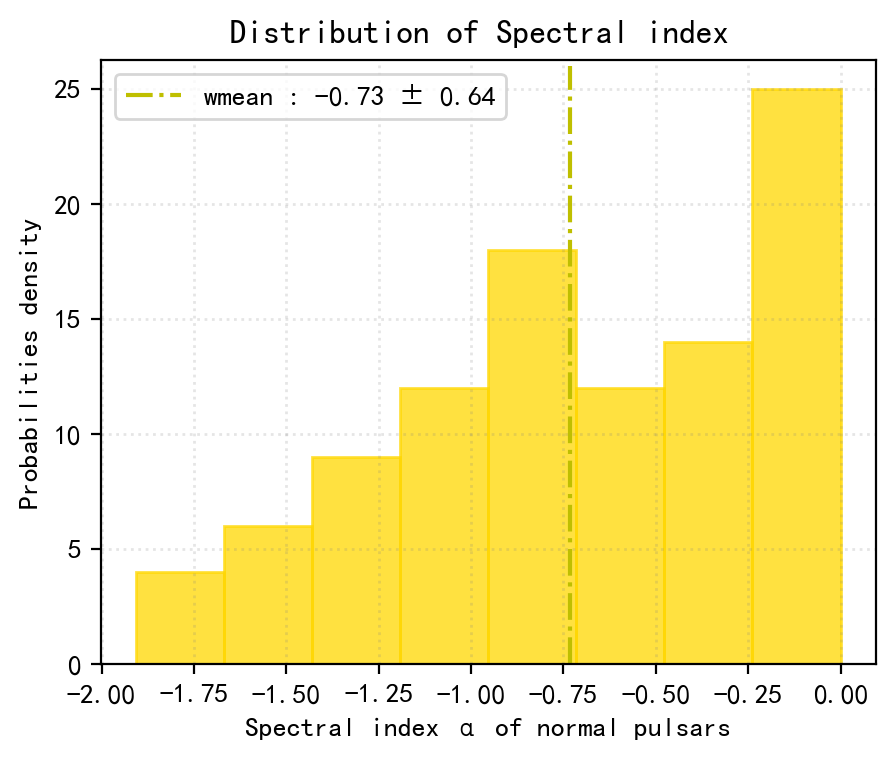}
    \caption{Histogram of the spectral indices $\alpha$ for normal pulsars that were classified to have high-frequency cut-off. The dashed line represents the weighted average.}
    \label{fig:3.1.4.1}
\end{figure}

Of the 800 normal pulsars in the sample, $12.50\%$ are part of the high-frequency cut-off model, which is the most common model apart from the simple power-law model. A histogram of the spectral index shown in Figure \ref{fig:3.1.4.1} revealed an average weight of $-0.73\pm0.64$4, with a range from $-2$ to $0$.

\begin{figure}
	\includegraphics[width=\columnwidth]{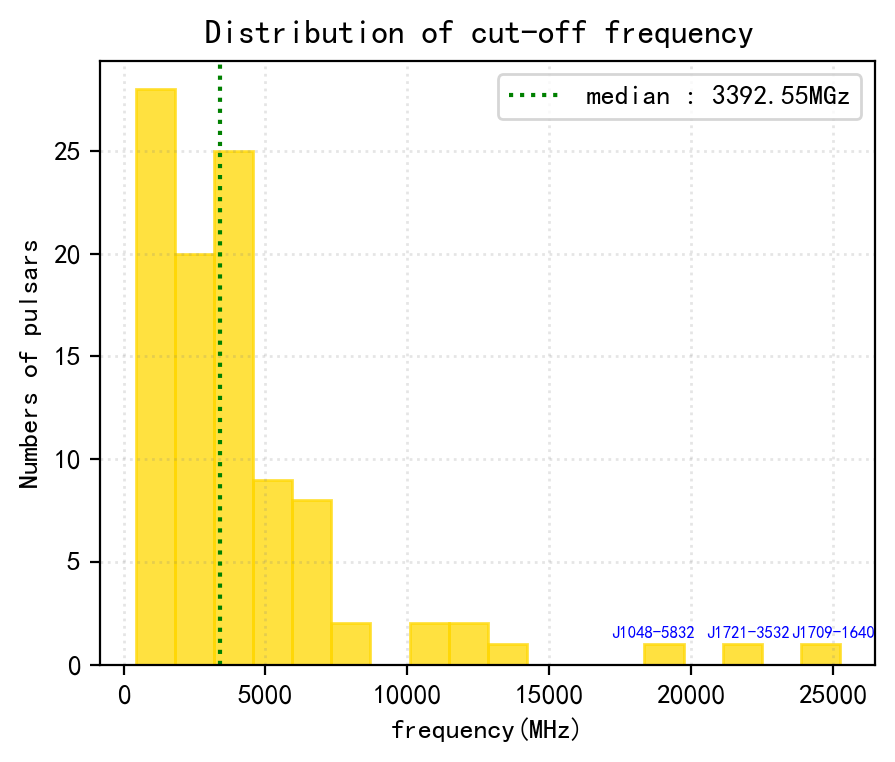}
    \caption{Distribution of cut-off frequency. We have marked the position of the median with a dashed line in the figure, and provided its value. \textbf{The three pulsars exhibiting the largest cut-off frequencies are identified as J1048-5832, J1721-3532, and J1709-1640.}}
    \label{fig:3.1.4.2}
\end{figure}

The cut-off frequency of pulsars is generally below $15$ GHz, as shown in Figure \ref{fig:3.1.4.2}. However, there are exceptions like PSRs J1048--5832, J1721--3532, and J1709-1640, with J1709-1640 having the highest cut-off frequency of $25.3$ GHz. The graph clearly indicates a higher concentration below 10 GHz for the high-frequency cut-off, with the median position at 3392 MHz. \textbf{We observe that three pulsars with the largest cut-off frequencies exhibit characteristics indicative of youth and stronger magnetic field intensities compared to the majority of pulsars. Specifically, of the three pulsars examined in this study, the youngest is PSR J1048-5832, estimated to be 20,400 years old, making it the second youngest pulsar in the model. Even the oldest pulsar, PSR J1709-1640 with an age of merely 1.64 million years, is younger than more than half of the pulsars included in this model. They posses larger magnetic field strength compared to approximately seventy percent of the pulsars in our sample.}

\subsubsection{Double-turn-over spectrum model}

\begin{figure}
	\includegraphics[width=\columnwidth]{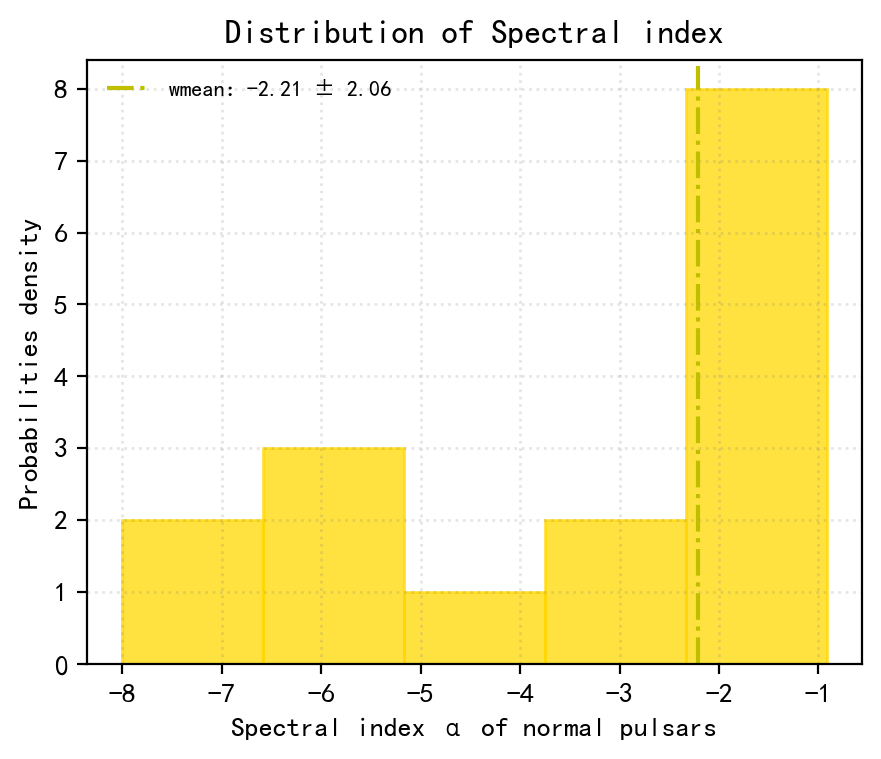}
    \caption{Histogram of the spectral indices $\alpha$ for normal pulsars that were classified to have \textbf{Double-turn-over spectrum.} The dashed line represents the weighted average.}
    \label{fig:3.1.5.1}
\end{figure}

The double-turn-over model simultaneously includes a low-frequency turn-over and a high-frequency cut-off phenomena. This particular model has the smallest correlation coefficient among the five models, just making up approximately $2.00\%$ of the total. They exhibit a weighted average $-2.21 \pm 2.06$ of spectral index as observed from the spectral index in Figure \ref{fig:3.1.5.1}.

\begin{figure}
	\includegraphics[width=\columnwidth]{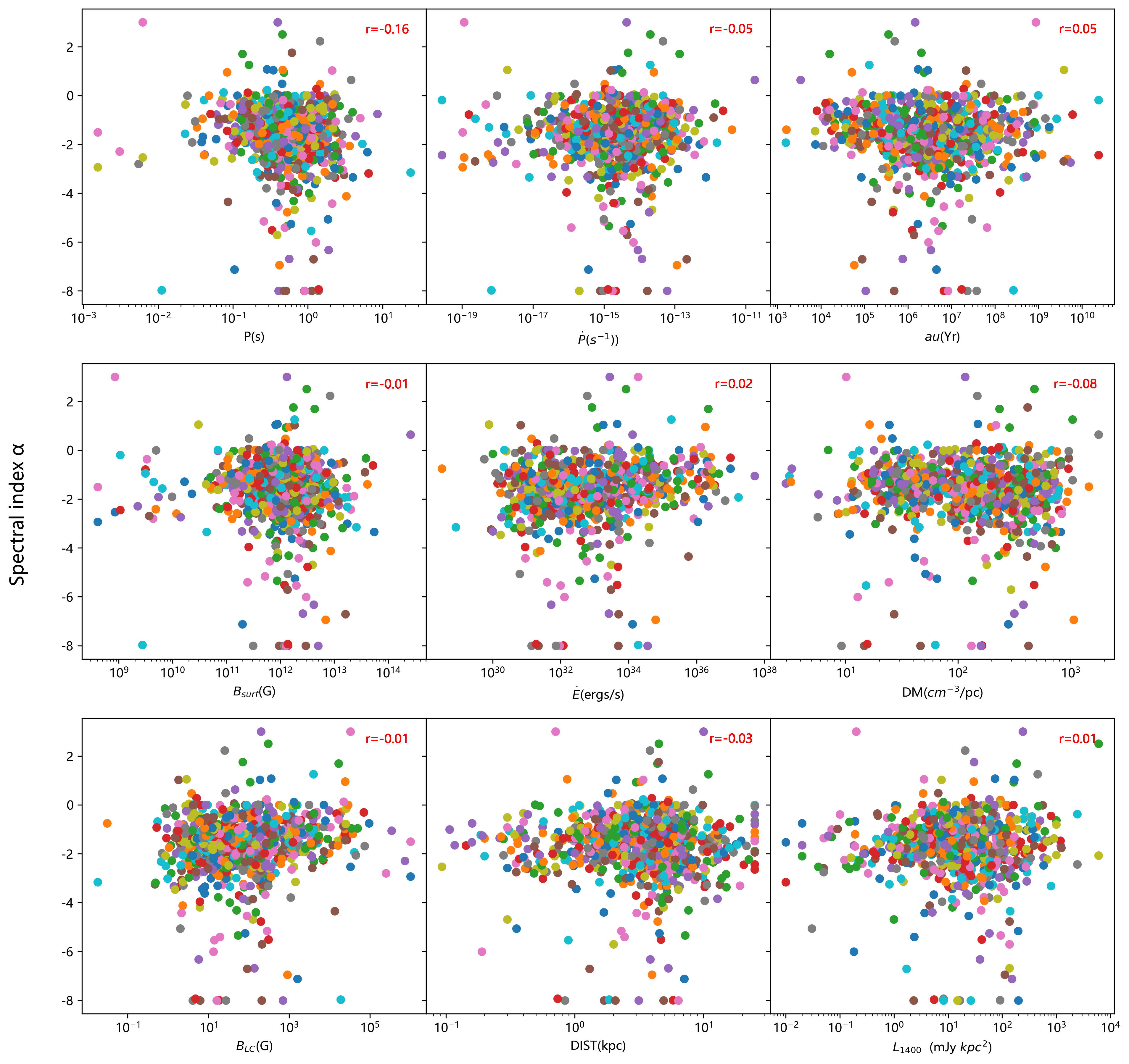}
    \caption{Distributions of spectral index $\alpha$ of normal pulsars vs pulsar period $P$, period derivative $\dot{P}$, characteristic age $\tau$, 1400 MHz pseudo-luminosity $L_{\rm {1400}}$, dispersion measure (DM), surface magnetic field $B_{\rm {\text{surf}}}$, the pulsar distance (DIST), Magnetic field at light cylinder $B_{\rm {\text{LC}}}$, spin-down luminosity $\dot{E}$.. The parameters of the fit and the Spearman rank coefficient are given at the top right corner of each subplot.}
    \label{fig:3.1.5.2}
\end{figure}

In addition, we separately discussed the correlation between spectral indices and characteristic parameters for each model. The results shown in Table \ref{tab:table4} indicate that there is no strong correlation between spectral indices and characteristic parameters by means of any model. We also analyzed the correlation between the spectral index and the characteristic parameters of all 
 normal pulsars. Specifically, for the broken power-law model, we separated the spectral index into two parts: before and after the break. We then considered them as separate spectral indices for two individual power-law models. Figure \ref{fig:3.1.5.2} presents the corresponding results.

\setcounter{table}{2}
\begin{table}
\caption{\textbf{ Correlations between the intrinsic parameters of normal pulsars and the spectral indices across various models, along with the corresponding $p_{\rm values}$. In this study, we find no significant correlation between the spectral index of normal pulsars and their characteristic parameters, both at the overall level and in terms of individual models. }} 
\centering 
\begin{threeparttable}
\label{tab:table4}
\setlength{\tabcolsep}{3mm}
\begin{tabular}{lccccccc}
\toprule 
Parameters\tnote{1} &sim  & \multicolumn{2}{c}{bro} & dou   & low   & high  & Total \tnote{2} \\
\midrule 
P                 & -0.20 (0.00)  & -0.04 (0.80)      & -0.42 (0.00)      & -0.18 (0.50) & -0.11 (0.32)& -0.06 (0.54) & -0.16 (0.00)\\
$\dot{P}$         & -0.07 (0.10) & 0.10 (0.52)       & -0.01 (0.97)     & 0.09 (0.75) & 0.14  (0.21)& -0.01 (0.90) & -0.05 (0.14)\\
$\tau$            & 0.05 (0.22)  & -0.10 (0.52)      & -0.15 (0.32)     & -0.23 (0.40)& -0.13 (0.23)& 0  (0.99)&0.05 (0.17)\\
$B_{\rm {\text{surf}}}$ & 0 (0.97)     & 0.06 (0.71)      & -0.16 (0.29)     & 0.03 (0.91) & 0.07  (0.51)& -0.06 (0.56) &-0.01 (0.74)\\
$\dot{E}$         & 0.05 (0.26)  & -0.18 (0.23)     & 0.10 (0.53)      & -0.25 (0.35)& 0.01  (0.93)& 0.09 (0.38)  &0.02 (0.54)\\
DM                & -0.10 (0.02)  & 0.16 (0.30)      & -0.11 (0.49)     & -0.21 (0.43)& -0.14 (0.21)& 0.07 (0.48)  &-0.05 (0.03)\\
$B_{\rm {\text{LC}}}$   & -0.03 (0.50) & 0.06 (0.71)      & 0.33 (0.03)      & 0.17  (0.52)& 0.09  (0.42)& 0.01  (0.56)&-0.01 (0.83)\\
Dist              & -0.09 (0.03) & 0.11 (0.46)      & -0.10 (0.51)     & -0.49 (0.05)& -0.17 (0.12)& 0.15  (0.15)&-0.03 (0.33)\\
$L_{\rm {1400}}$        & 0.07 (0.10)  & 0.13 (0.38)      & 0.04 (0.80)      & -0.11 (0.68)& -0.04 (0.73)& -0.18 (0.07)&0.01 (0.74)\\
\bottomrule 
\end{tabular}
\begin{tablenotes}
\item[1] Pulsar characteristic parameters : pulsar period $P$, period derivative $\dot{P}$, characteristic age $\tau$, 1400 MHz pseudo-luminosity $L_{\rm {1400}}$, dispersion measure (DM), surface magnetic field $B_{\rm {\text{surf}}}$, the pulsar distance (DIST), Magnetic field at light cylinder $B_{\rm {\text{LC}}}$, spin-down luminosity $\dot{E}$. 
\item[2] Here, abbreviations are used for the names of the models, and the names after the abbreviations are indicated in parentheses: the simple power law(sim), broken power law(bro), double tur-nover spectrum(dou), low-frequencies turn-over(low) and high-frequencies cut-off(high). The last column represents all normal pulsars without distinguishing the models.
\end{tablenotes}
\end{threeparttable}
\end{table}

\subsection{Millisecond pulsars(MSP)}

\setcounter{table}{3}
\begin{table*}
\caption{\textbf{Correlations between the intrinsic parameters of millisecond pulsars (MSPs) and the spectral indices across various models, along with the corresponding $p_{\rm values}$. In line with normal pulsars, no significant correlation is observed between the spectral index and the characteristic parameters at the overall level. However, for individual models, with the exception of the simple power-law model, a noteworthy correlation is observed between the spectral index and certain characteristic parameters. To emphasize this finding, we have highlighted the relevant coefficients. Bold font indicates that the absolute value of correlation is greater than 0.5.}}
\centering 
\begin{threeparttable}
\label{tab:table3}
\setlength{\tabcolsep}{3mm}
\begin{tabular}{lccccccc} 
\toprule 
Parameters\tnote{1} &sim  & \multicolumn{2}{c}{bro} & dou   & low   & high  & Total \tnote{2} \\
\midrule 
P                 & 0.12 (0.39) & \textbf{-0.5} (0.17)     & \textbf{-0.52} (0.15)     & \textbf{0.75} (0.05) & 0.15 (0.70) & -0.14 (0.76)&0.03 (0.75)\\
$\dot{P}$         & 0 (0.99)   & 0.05 (0.90)      & -0.12 (0.77)     & \textbf{0.71} (0.07) & 0.48 (0.19) & 0.36 (0.43) &0.06 (0.56)\\
$\tau$            & -0.24 (0.08)& -0.07 (0.86)     & -0.37 (0.33)     & \textbf{-0.71} (0.07)& -0.26 (0.50)& \textbf{-0.71} (0.07)&0.24 (0.02)\\
$B_{\rm {\text{surf}}}$ & 0.02 (0.88) & -0.13 (0.75)     & -0.19 (0.62)     & -0.43 (0.34)& \textbf{-0.68} (0.04)& -0.14 (0.76)&-0.07 (0.50)\\
$\dot{E}$         & -0.09 (0.51)& 0.40 (0.29)       & \textbf{-0.60} (0.09)     & \textbf{-0.61} (0.15)& -0.24 (0.53)& 0.46 (0.29) &-0.11 (0.28)\\
DM                & -0.35 (0.01)& -0.18 (0.64)     & -0.32 (0.41)     & \textbf{-0.64} (0.12)& 0.04 (0.91)& 0.39 (0.38) &-0.26 (0.01)\\
$B_{\rm {\text{LC}}}$   & -0.12 (0.39)& \textbf{0.5} (0.17)       & \textbf{0.57} (0.11)      & -0.07 (0.88)& -0.46 (0.21)& 0.32 (0.48)&-0.13 (0.23)\\
Dist              & -0.44 (0.00)& \textbf{0.6} (0.09)       & 0.02 (0.97)      & \textbf{-0.57} (0.18)& \textbf{-0.61} (0.08)& \textbf{0.57} (0.18)&-0.31 (0.00)\\
$L_{\rm {1400}}$        & -0.11 (0.43)& 0.02 (0.97)      & 0.48 (0.19)      & 0 (1.00)   & -0.27 (0.49)& \textbf{0.79} (0.04)&0.00 (0.97)\\
\bottomrule 
\end{tabular}
\begin{tablenotes}
\item[1] Pulsar characteristic parameters : pulsar period $P$, period derivative $\dot{P}$, characteristic age $\tau$, 1400 MHz pseudo-luminosity $L_{\rm {1400}}$, dispersion measure (DM), surface magnetic field $B_{\rm {\text{surf}}}$, the pulsar distance (DIST), Magnetic field at light cylinder $B_{\rm {\text{LC}}}$, spin-down luminosity $\dot{E}$. 
\item[2] Here, abbreviations are used for the names of the models, and the names after the abbreviations are indicated in parentheses: the simple power law(sim), broken power law(bro), double turn-over spectrum(dou), low-frequencies turn-over(low) and high-frequencies cut-off(high). The last column represents all MSPs without distinguishing the models.
\end{tablenotes}
\end{threeparttable}
\end{table*}

The criteria for identifying millisecond pulsars are established in \cite{lorimer2008}. It is specified so that millisecond pulsars should have periods ranging from $1.4$ ms to $30$ ms, and their period derivatives should be less than $10^{-19}$ s/s. \textbf{Based on these criteria, we selected 86 MSPs, whose spectral behavior was also described with the five models employed above}. Among these pulsars, $62.79\%$ can be adequately represented using a simple power law model. The remaining millisecond pulsars exhibit a nearly uniform distribution among the other models. As shown in Table \ref{tab:table00}, the correlation coefficient of the power-law model and the low-frequency turn-over is approximately 10.47\%, while the correlation coefficients of the high-frequency cut-off and double turn-over model are both  8.14\%.

\subsubsection{Simple power-law model}

The overall trend of the millisecond pulsar spectral index remains consistent with a normal distribution, with the weighted average of $-1.89 \pm 0.21$. For millisecond pulsars other than $J1802-2124$ ($0.042$), $J2017+0603$ ($1.036$), and $J2215+5135$ ($-3.702$), the spectral index generally ranges from $-3$ to $0$.%, as shown in Figure \ref{fig:3.2.1.1}.

%\begin{figure}
	%\includegraphics[width=\columnwidth]{3.2.1.1.png}
   % \caption{Histogram of the spectral indices $\alpha$ for MSPs that were classified to have Simple %power-law.}
   % \label{fig:3.2.1.1}
%\end{figure}

%Finally, we performed a Spearman analysis to examine the correlation between MSPs following a simple power law and the characteristic parameters of pulsars. The results in the Figure \ref{fig:3.2.2.2} indicated that, apart from the spectral index, which displayed correlation coefficients of $-0.44$ and $-0.35$ with respect to the pulsar distance and dispersion measure (DM), respectively, the remaining correlations were not sufficiently strong.

\subsubsection{Broken power-law spectra model}
\label{subsec:bplmp}

There are 9 millisecond pulsars (MSPs) that conform to the broken power-law model. The millisecond pulsars' spectral indices before the break range from $-1.79$ to $0.937$, with a weighted average of $-1.12$. After the break, the spectral indices are distributed between $-2.827$ and $-0.007$, with a weighted average of $-2.48$. However, for J1918-0642, the spectral index reaches $-8$ after the break, which is in close proximity to our predefined lower fit limit.

When discussing the correlation between the spectral index and characteristic parameters, millisecond pulsars exhibit broader differences compared to the normal pulsars in Table \ref{tab:table3}.  Notably, the correlation between the spectral index before the break and the pulsar distance represents a value of $0.60$. Similarly, the correlation between the spectral index after the break and the spin-down energy loss rate also indicates a value of $0.60$.

\subsubsection{Power-law spectra with low-frequency turn-over}

Additionally, there are 9 millisecond pulsars (MSPs) that conform to the low-frequency turn-over model. Overall, the weighted average spectral index is significantly low, measuring $-5.54$. J$0024-7204C$ and J$0024-7204$ exhibit a spectral index of $-8$, whereas J$1600-3053$ has a spectral index of $-7.572$. Based on the frequency distribution, it is evident that the breaks exhibit a notable concentration within the $100$ MHz to $1$ GHz range.

Moreover, we have examined the correlations between the spectral index and characteristic parameters of pulsars. The results reveal a strong negative correlation of $-0.68$ between the spectral index and the intensity of the magnetic field, as well as a negative correlation of $-0.61$ between the spectral index and the distance of the pulsars. Moreover, we noticed a positive correlation of $0.48$ between the spectral index and the period derivative, and a negative correlation of $-0.46$ between the spectral index and the magnetic field intensity at the light cylinder (see Table \ref{tab:table3}). The results may be an indicative of evolution of the millisecond pulsars in low mass X-ray binaries (LMXBs). In the Roche lobe overflow phase of LMXBs, matter carrying high specific angular momentum from a low mass companion star is transferred onto the neutron star through an accretion disk, leading to acceleration of rotation of the underlying neutron star. Also settled matter in the form of highly conductive plasma causes magnetic field to decay by heating up the surface of the neutron star \citep{Shibazaki1989} and shielding the field lines via diamagnetic screening effect \citep{Choudhuri2004}. If this binary evolution stage lasts for $\lesssim$\,Gyr, that neutron star ends up with a millisecond pulsar attaining fast rotation close to the maximum possible equilibrium period with a residual magnetic field. As a consequence of such evolution, the millisecond pulsars possess a narrow range both for $P$, $\dot{P}$ and $B_{\rm LC}$ compared to the radio pulsar population, which may be the underlying reason for the corresponding correlations.

\subsubsection{Power-law spectra with high-frequency cut-off}

A total of 7 millisecond pulsars have been included in the high-frequency cut-off model. The distribution of their spectral indices is concentrated, ranging between $-1.374$ and $0$, with a weighted average spectral index of $-1.02$. Looking at the cut-off positions, except J$1810+1744$ at $416$ MHz, all millisecond pulsars have cut-off frequencies above $1.5$ GHz. Additionally, under this spectra fitting model, the correlation coefficients between millisecond pulsars' spectral index and characteristic age and radio luminosity at $1400$ MHz, are $-0.71$ and $0.79$, respectively. The correlation coefficient between millisecond pulsars' spectral index and their distance reaches $0.57$ (see Table \ref{tab:table3}).

\subsubsection{Double-turn-over spectrum model}

Seven millisecond pulsars conforming to the double reversal model were identified, all with spectral indices below $0$ and a weighted average $-2.82 \pm 2.58$. Among these pulsars, $J1911-1114$ exhibits the smallest spectral index of $-8$. With the exception of $J0024-7204D$ at $689.5$ MHz, the peak frequencies of these pulsars are all below $200$ MHz, while the cut-off positions for the spectral indices are above $4$ GHz.

Moreover, a closer examination revealed strong correlations between spectral indices and characteristic parameters (see Table \ref{tab:table3}), including period, period derivative, and characteristic age, with correlation coefficients of $0.75$, $0.71$, and $-0.71$. Additionally, the spectral index shows significant correlations with spin-down power ($-0.61$), dispersion measure ($0.64$), and pulsar distance ($-0.57$).

\begin{figure}
	\includegraphics[width=\columnwidth]{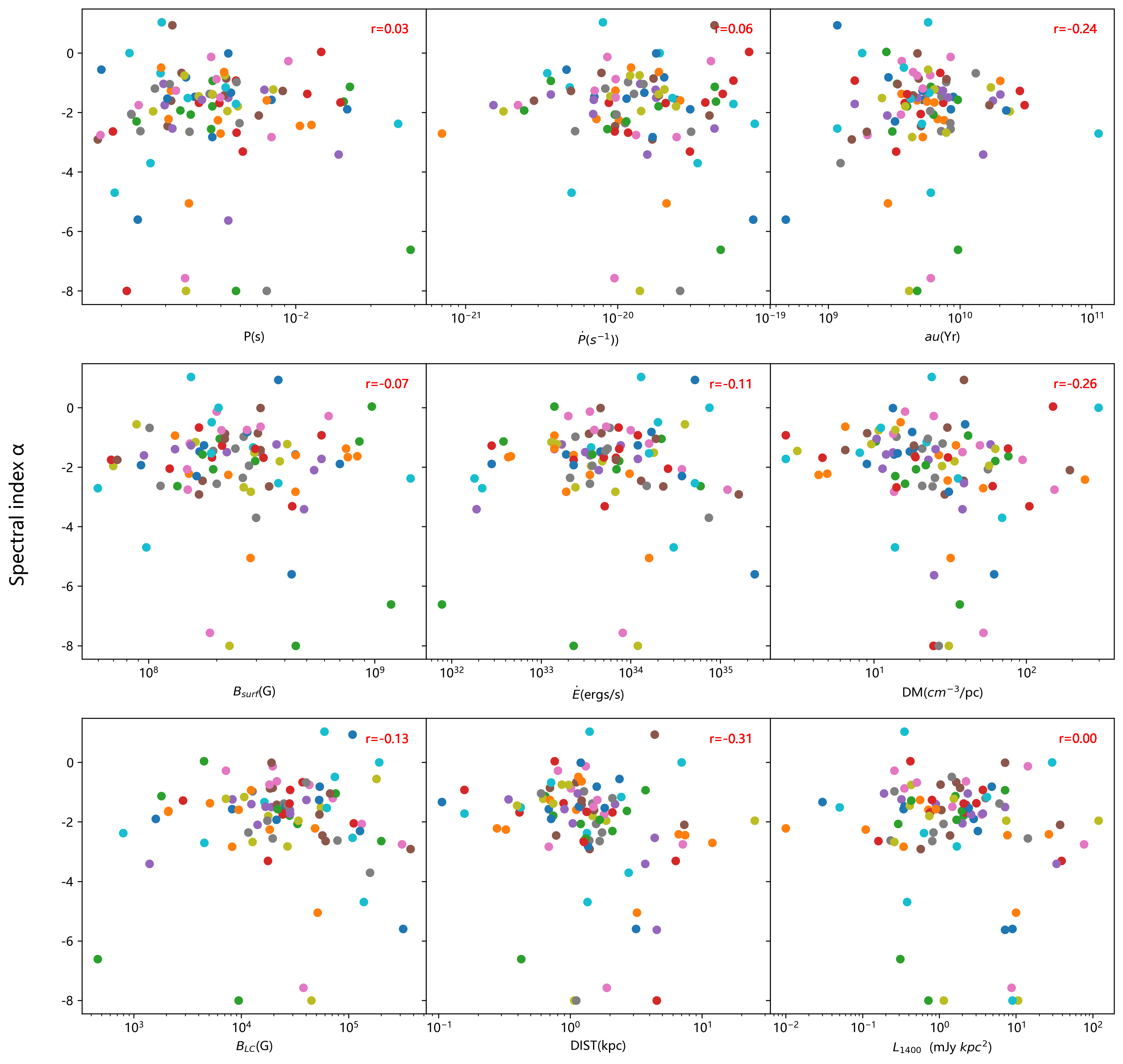}
    \caption{Distributions of spectral index $\alpha$ of MPSs vs pulsar period $P$, period derivative $\dot{P}$, characteristic age $\tau$, 1400 MHz pseudo-luminosity $L_{\rm {1400}}$, dispersion measure (DM), surface magnetic field $B_{\rm {\text{surf}}}$, the pulsar distance (DIST), Magnetic field at light cylinder $B_{\rm {\text{LC}}}$, spin-down luminosity $\dot{E}$. The parameters of the fit and the Spearman rank coefficient are given at the top right corner of each subplot.}
    \label{fig:3.2.5.1}
\end{figure}

Finally, we analyzed the correlation between the spectral index and the characteristic parameters of all millisecond pulsars, just like in the previous section. The final results are shown in Figure \ref{fig:3.2.5.1}. Overall, there is not a strong correlation observed.

\subsection{The gigahertz-peaked spectrum(GPS)}
\label{sec:gps}

The criteria set by \citet{jankowski2018} was used to identify GPS pulsars, whereby pulsars displaying peak frequencies between $0.6$ and $2$ GHz were selected, without considering the best-fit models. From our analysis, 33 GPS pulsars meet these criteria, with only $11$ previously identified (see Table \ref{tab:table1}). Notably, of the newly identified GPS pulsars, $4$ showed spectral breaks, $16$ exhibited the low-frequency turn-over feature, while $2$ appeared in double-turn-over models.

We compared our results with those of \citet{kijak2021low} and discovered that $22$ pulsars did not satisfy our GPS criteria. Specifically, only $4$ GPS pulsars have flux density values at three frequencies, thereby failing to fulfill our requirement that all the pulsars we are studying must provide four or more flux density values at different frequencies. Five of the pulsars exhibited peak frequencies below $600$ MHz, while seven others obeyed to a simple power-law model, and the remaining six utilized high-frequency cut-off models. We excluded these models from our GPS pulsars because they did not conform with our "peak" criteria.

Statistical analysis was performed on the confirmed 33 GPS pulsars, of which only approximately two-thirds exhibited high dispersion measures \textbf{($\text{DM} > 150 \, \text{cm}^{-3} \, \mathrm{pc}$)}. We also investigated the correlation between $v_{\text{peak}}$ and various characteristic parameters of the pulsars. The correlation with DM was found to be weak ($r = 0.20$), providing further support for the hypothesis proposed by \citep{kijak2007} that the occurrence of higher DM values in GPS pulsars is due to selection effects. Additionally, the results revealed a significant correlation between $v_{\text{peak}}$ and both the period ($r = 0.40$) and the period derivative ($r = 0.49$).  Furthermore, a weak negative correlation was observed between $v_{\text{peak}}$ and the characteristic age ($r = -0.05$).

\setcounter{table}{4}
\begin{table*}
\caption{\textbf{Some parameters of GPS pulsars, where $\nu_{p}$ and $\nu_{p}$\_err are the the peak/turn-over frequency and its error, $P$ is the pulsar period , $\dot{P}$ is the period derivative, $\tau$ is the characteristic age, DM is the dispersion measure.}} 
\centering 
\begin{threeparttable}
\label{tab:table1}
\setlength{\tabcolsep}{2mm}
\begin{tabular}{lccccccc} 
\toprule 
PSRJ\tnote{1}  & model\tnote{2} & $\nu_{p}*$ (MHz)  & $\nu_{p}$\_err (MHz)  & P (s)    & $\dot{P}$ (s/s$^{-1}$)   & DM (cm$^{-3}$ pc) & $\tau$ (yr)    \\
\midrule 
J0024-7204C & low   & 600.4  & 9          & 5.76E-03 & -4.99E-20   & 24.5955                       & *          \\
J0024-7204D & dou   & 689.5  & 40.6       & 5.36E-03 & -3.42E-21   & 24.7432                       & *          \\
J0024-7204J & low   & 607.8  & 4.5        & 2.10E-03 & -9.79E-21   & 24.5932                       & *          \\
J0908-4913  & low   & 603.5  & 42.3       & 1.07E-01 & 1.51E-14    & 180.44                        & 112000     \\
J1326-5859  & bro   & 647.1  & 201.7      & 4.78E-01 & 3.2418E-15  & 287.17                        & 2340000    \\
J1513-5908  & low   & 611.6  & 65.6       & 1.52E-01 & 1.53E-12    & 252.5                         & 1570       \\
J1524-5706  & low   & 1189.4 & 65.2       & 1.12E+00 & 3.56E-13    & 832                           & 49600      \\
J1600-3053  & low   & 847    & 18.7       & 3.60E-03 & 9.50E-21    & 52.3282                       & 6000000000 \\
J1635-5954  & low   & 857.2  & 59.9       & 5.29E-01 & 1.37E-15    & 134.9                         & 6130000    \\
J1644-4559  & bro   & 729.9  & 3          & 4.55E-01 & 2.00892E-14 & 478.66                        & 359000     \\
J1705-3950  & low   & 1039.4 & 71.9       & 3.19E-01 & 6.06E-14    & 207.25                        & 83400      \\
J1727-2739  & bro   & 939.9  & 231.6      & 1.29E+00 & 1.07E-15    & 146                           & 19100000   \\
J1740-3015  & low   & 951.4  & 146.7      & 6.07E-01 & 4.66E-13    & 151.96                        & 20600      \\
J1741-3016*  & dou   & 947.9  & 26.3       & 1.89E+00 & 8.99E-15    & 382                           & 3340000    \\
J1745-3040  & low   & 1009.9 & 130.2      & 3.67E-01 & 1.07E-14    & 88.373                        & 546000     \\
J1751-3323  & low   & 1006.2 & 109.3      & 5.48E-01 & 8.83E-15    & 296.7                         & 984000     \\
J1753-2501  & low   & 760.8  & 91.1       & 5.28E-01 & 1.41E-14    & 672                           & 593000     \\
J1803-2137  & bro   & 835.2  & 33.5       & 1.34E-01 & 1.34359E-13 & 233.99                        & 15800      \\
J1806-1154* & low   & 642.3  & 64.7       & 5.23E-01 & 1.41E-15    & 122.41                        & 5880000    \\
J1809-1917  & bro   & 1908.2 & 231.1      & 8.28E-02 & 2.55302E-14 & 197.1                         & 51400      \\
J1818-1422* & low   & 886.1  & 54         & 2.91E-01 & 2.04E-15    & 619.65                        & 2270000    \\
J1823-1115* & bro   & 628.4  & 56.1       & 2.80E-01 & 1.37862E-15 & 428.59                        & 3220000    \\
J1825-1446* & dou   & 1722.7 & 223.8      & 2.79E-01 & 2.27E-14    & 352.23                        & 195000     \\
J1826-1334* & low   & 1220.1 & 134.6      & 1.01E-01 & 7.53E-14    & 231                           & 21400      \\
J1830-1059* & low   & 913.1  & 47.8       & 4.05E-01 & 5.99E-14    & 159.7                         & 107000     \\
J1832-1021* & dou   & 2000   & 1483.6     & 3.30E-01 & 4.20E-15    & 474.14                        & 1250000    \\
J1843-0459* & low   & 807.4  & 59.1       & 7.55E-01 & 8.54E-16    & 444.1                         & 14000000   \\
J1844-0244* & low   & 861.1  & 45.7       & 5.08E-01 & 1.67E-14    & 422.13                        & 481000     \\
J1844-0538* & low   & 1008   & 57.2       & 2.56E-01 & 9.71E-15    & 411.71                        & 417000     \\
J1845-0743* & low   & 804.5  & 9.1        & 1.05E-01 & 3.67E-16    & 280.93                        & 4520000    \\
J1852-0635* & low   & 1119   & 41.1       & 5.24E-01 & 1.46E-14    & 173.9                         & 568000     \\
J1918+1444* & low   & 1998.4 & 1640.8     & 1.18E+00 & 2.12E-13    & 27.202                        & 88100      \\
J2321+6024  & bro   & 1161.7 & 66.9       & 2.26E+00 & 7.03689E-15 & 94.591                        & 5080000  \\
\bottomrule 
\end{tabular}
\begin{tablenotes}
\item[1] The asterisk (*) represents GPS pulsars identified in previous literature.
\item[2] Here, abbreviations are used for the names of the models, and the names after the abbreviations are indicated in parentheses: broken power law(bro), double turn-over spectrum(dou) and low-frequencies turn-over(low)
\end{tablenotes}
\end{threeparttable}
\end{table*}

\section{DISCUSSION}
\label{sec:discussion}

Recently, there has been extensive research investigating the determination of flux and spectral indices of pulsars, nonetheless, there is still no definite explanation for the radiation mechanism and spectral turn-over phenomenon of pulsars. To understand these features better, it is essential to investigate pulsar flux density and the corresponding spectrum. Since the frequency dependence of pulsar emission is a crucial characteristics, the spectrum is intimately related to the radiative mechanism of pulsars. Such an understanding is especially crucial for examining the physical conditions in the magnetosphere. Therefore, the analysis of pulsar flux density and its spectral indices is essential for providing further insights into pulsars. Through sampling a larger group, we aim to investigate the similarities and differences in spectral indices between millisecond pulsars and normal pulsars.

\subsection{Spectral indices of Normal pulsars and millisecond pulsars}

Several reports on the average spectral indices of pulsars have been generated from the previous spectral studies. 
\citet{Sieber1973} reported an average spectral index of approximately $-1.62$ after examining the spectral indices of $27$ pulsars known to that date. \textbf{\citet{kramer1998} suggested that the average spectral index of millisecond pulsars (MSPs) should be slightly steeper than that of ordinary pulsars, with being approximately $-1.8$.} This value remains consistent with the average value of $-1.8 \pm 0.2$ from the simple power-law fit reported by \citet{maron2000}. Additionally, \citet{Malofeev1994} noted that pulsars with spectral turn-over typically demonstrate a low-frequency turn-over at approximately $0.1$ GHz and a high-frequency cut-off frequency ranging from $0.4$ to $9.1$ GHz. These pulsars exhibit a steeper spectral index after the turn-over. The results from analysis our sample suggest that the spectral index shows a steeper trend after a spectral break in the broken power-law model, regardless of whether they are normal pulsars or MSPs. \textbf{In a recent study, \citet{Spiewak2022} reported an average spectral index of approximately $-1.9$ for $189$ MSPs. More recently, a more specific study of flux densities and spectral indices of MSPs with MeerKAT has been done by \citet{gitika2023flux}. They reported a mean spectral index of $-1.86(6)$ for $89$ MSPs. This value appears to be close to the previous results.} Our simple power-law model produced a similar average spectral index of approximately $-1.61$, which is comparable to the reported average spectral index of $276$ pulsars by \citet{jankowski2018}. Further analysis reveals that only about $29$\% of our sample overlaps with \citet{jankowski2018}'s dataset. This indicates that the average spectral index is relatively independent to a large extent.

The weighted average of the spectral index from each model reveals that, except for the broken power-law model, millisecond pulsars exhibit steeper spectral index compared to normal pulsars in the other models in Table \ref{tab:table2}. The results here are consistent with those reported in \cite{lorimer1995}, where millisecond pulsars exhibit steeper spectral indices than normal pulsars. They argue that this may be caused by the characteristic age of pulsars. As is shown in the subsequent section, an analysis of the correlation between the spectral index and characteristic ages of millisecond pulsars revealed no significant conformity. A comparison of the different models reveals that both millisecond pulsars and normal pulsars exhibit the lowest spectral index in the low-frequency turn-over model, followed by the double turn-over model, the simple power-law model, and finally the high-frequency cut-off model. However, in the case of the broken power-law model, if we consider only the spectral index before the spectral break, it emerges out as the second highest. This phenomenon could arise from various spectral behavior patterns or other factors that influence the spectral index. These factors may include specific characteristics of pulsars, such as their period and characteristic age.

\setcounter{table}{5}
\begin{table}
\caption{\textbf{Weighted average of normal pulsars and millisecond pulsars under different models.}} 
\centering 
\begin{threeparttable}
\label{tab:table2}
\setlength{\tabcolsep}{4mm}
\begin{tabular}{lcccc} 
\toprule 
model\tnote{1} & Normal pulsars   & MSPs \\
\midrule 
simple power-law                       & -1.57±0.32               & -1.65±0.81 \\
broken power-law  \tnote{2}                     &  \makecell{-1.16±1.16 \\ -2.00±0.38}&  \makecell{-1.12±0.65 \\ -1.75±0.38}\\
low-frequency turn-over & -3.45±1.03               & -5.54±1.80  \\
high-frequency cut-off  & -0.73±0.64               & -1.02±0.17  \\
double-turn-over spectrum              & -2.21±2.06               & -2.82±2.58 \\
\bottomrule 
\end{tabular}
\begin{tablenotes}
\item[1]We calculated a weighted average by weighting each spectral index with its reciprocal error, and using the covariance matrix of the spectral index and its errors to determine the uncertainty. 
\item[2]The top line represents the spectrum before the break, while bottom line represents the spectral index after the break. 
\end{tablenotes}
\end{threeparttable}
\end{table}

\begin{figure}
	\includegraphics[width=\columnwidth]{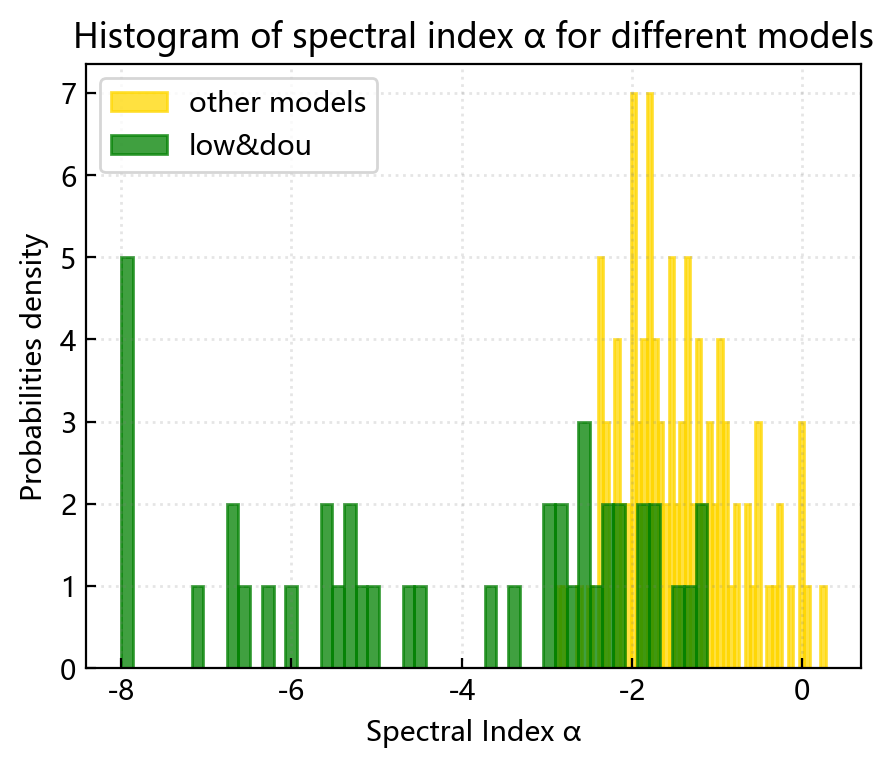}
   \caption{Histogram of the spectral indices $\alpha$ differenr models. The models are categorized into two groups: the low-frequency turnover and the double-turn-over model, and the remaining models fall into the other category.}
    \label{fig:fre}
\end{figure}

\textbf{In addition, we observed the presence of pulsars exhibiting steep spectral indices in the low-frequency turnover and double-turn-over models. By considering the most extreme spectral index value of $-4.42$ from the simple power-law model as a benchmark, we specifically selected $33$ pulsars with spectral indices steeper than this value, which exclusively exist in the low-frequency turnover and double-turn-over models. We then conducted a search for other spectral indices within the same frequency range of these identified pulsars. Given the varied frequency coverage, we focused solely on identifying spectral indices that completely matched at the same frequency, resulting in some pulsars out of the initial $33$ lacking matches in Table \ref{tab:tablefre}. Then, we plotted the spectral index distribution of these pulsars (See in Figure \ref{fig:fre}). Through comparison, we discovered that even within the same frequency range, pulsars exhibit diverse spectral shapes, characterized by different spectral indices and models. Notably, pulsars with extremely steep spectral indices do not demonstrate any other distinctive characteristics.}

\setcounter{table}{6}

\begin{longtable}{lcccc}
\caption{\textbf{The spectral indices within the same frequency range, bold are those with extremely steep spectral index values.}}  
\label{tab:tablefre}\\
\toprule
PSRJ\tnote{1} & $\alpha$ & $\alpha$\_err & Freq. range (MHz) & Model\tnote{2} \\
\midrule
\endfirsthead
\multicolumn{5}{c}{{\tablename} \thetable{} (Continued)} \\
\toprule
PSRJ\tnote{1} & $\alpha$ & $\alpha$\_err & Freq. range (MHz) & Model\tnote{2} \\
\midrule
\endhead
\bottomrule
\multicolumn{5}{r}{{Continued on the next page}} \\
\endfoot
\bottomrule
\endlastfoot
J0452-1759   & -0.34            & 0.07   & 102.5-4850 & bro   \\
J0621+1002*  & \textbf{-6.61}   & 0.12   & 102.5-4850 & low   \\
J0944-1354   & -1.82            & 0.08   & 102.5-4850 & sim   \\
J2337+6151   & -1.32            & 0.11   & 102.5-4850 & sim   \\
J1711-1509   & -1.73            & 0.09   & 102.5-4850 & sim   \\
J1623-0908   & -1.73            & 0.11   & 102.5-4850 & sim   \\
J1916+0951   & -1.86            & 0.06   & 102.5-4850 & sim   \\
J1823+0550   & -1.76            & 0.08   & 102.5-4850 & sim   \\
J0601-0527   & -1.77            & 0.10   & 102.5-4850 & low   \\
J0653+8051   & -1.57            & 0.10   & 102.5-4850 & sim   \\
J1748-1300   & -1.87            & 0.13   & 102.5-4850 & sim   \\
J0040+5716   & -0.96            & 0.51   & 102.5-4850 & bro   \\
J1834-0426   & -0.93            & 0.06   & 102.5-4850 & high  \\
J0758-1528   & -1.44            & 0.08   & 102.5-4850 & sim   \\
J2149+6329   & 0.06             & 0.19   & 102.5-4850 & bro   \\
J1910+0358   & -1.52            & 0.09   & 102.5-4850 & sim   \\
J0629+2415   & -1.38            & 0.08   & 102.5-4850 & sim   \\
J2150+5247   & -0.74            & 0.13   & 102.5-4850 & high  \\
J1913-0440   & -2.24            & 0.22   & 102.5-4850 & low   \\
J0406+6138   & \textbf{-5.26}   & 0.07   & 102.5-4850 & low   \\
J1926+0431   & -1.19            & 0.21   & 102.5-4850 & high  \\
J2055+3630   & -0.65            & 0.35   & 102.5-4850 & bro   \\
J1913+1400   & -1.53            & 0.08   & 102.5-4850 & sim   \\
J1754+5201   & -0.11            & 0.22   & 102.5-4850 & high  \\
J2002+4050   & \textbf{-8.00}   & 1.34   & 102.5-4850 & low   \\
J2002+3217   & -1.10            & 0.08   & 102.5-4850 & sim   \\
J1915+1009   & -1.94            & 0.09   & 102.5-4850 & sim   \\
J2013+3845   & -0.85            & 0.08   & 102.5-4850 & high  \\
J1857+0212   & -1.29            & 0.14   & 102.5-4850 & sim   \\
J1257-1027   & -1.47            & 0.08   & 102.5-4850 & sim   \\
J1116-4122   & -1.39            & 0.07   & 147.5-3100 & high  \\
J1045-4509*  & -1.24            & 0.04   & 147.5-3100 & high  \\
J0907-5157   & -1.21            & 0.11   & 147.5-3100 & low   \\
J0959-4809   & -1.59            & 0.09   & 147.5-3100 & sim   \\
J0855-3331   & -2.61            & 0.67   & 147.5-3100 & low   \\
J0846-3533   & -2.32            & 0.24   & 147.5-3100 & low   \\
J1751-4657   & -2.78            & 0.18   & 147.5-3100 & low   \\
J1823-3106   & -1.58            & 0.06   & 147.5-3100 & sim   \\
J1034-3224   & -1.19            & 0.07   & 147.5-3100 & sim   \\
J0857-4424   & -2.08            & 0.05   & 147.5-3100 & sim   \\
J1722-3207   & -2.50            & 0.36   & 147.5-3100 & low   \\
J1801-2920   & -2.15            & 0.45   & 147.5-3100 & low   \\
J1604-4909   & -0.93            & 0.16   & 147.5-3100 & bro   \\
J1743-3150   & -3.00            & 0.20   & 147.5-3100 & low   \\
J1613-4714   & -1.70            & 0.11   & 147.5-3100 & sim   \\
J1801-2451   & -0.28            & 0.15   & 147.5-3100 & high  \\
J1614-5048   & -2.00            & 0.06   & 147.5-3100 & sim   \\
J1836-1008   & \textbf{-6.69}   & 0.05   & 147.5-3100 & low   \\
J1824-2452A  & -2.29            & 0.02   & 147.5-3100 & sim   \\
J1633-5015   & -1.04            & 0.10   & 147.5-3100 & bro   \\
J1717-3425   & -2.43            & 0.24   & 147.5-3100 & low   \\
J1559-4438   & \textbf{-5.16}   & 0.01   & 147.5-5124 & low   \\
J1651-4246   & -2.16            & 0.01   & 147.5-5124 & bro   \\
J1703-3241   & -1.34            & 0.07   & 147.5-5124 & high  \\
J0856-6137   & -2.20            & 0.08   & 151.5-3100 & sim   \\
J0840-5332   & -3.02            & 2.18   & 151.5-3100 & low   \\
J1001-5507   & -0.51            & 0.52   & 151.5-3100 & bro   \\
J2053-7200   & -1.16            & 0.19   & 151.5-3100 & high  \\
J1121-5444   & -1.74            & 0.10   & 151.5-3100 & high  \\
J0942-5657   & -2.15            & 0.11   & 151.5-3100 & sim   \\
J0924-5302   & -1.98            & 0.07   & 151.5-3100 & sim   \\
J0942-5552   & \textbf{-4.68}   & 0.03   & 151.5-3100 & low   \\
J0905-5127   & -1.84            & 0.09   & 151.5-3100 & sim   \\
J1057-5226   & -2.53            & 0.46   & 151.5-3100 & low   \\
J0953+0755   & -1.36            & 0.02   & 20-22700   & dou   \\
J0814+7429   & -1.81            & 0.05   & 20-22700   & low   \\
J1239+2453   & \textbf{-8.00}   & 5.10   & 20-22700   & dou   \\
J0826+2637   & -1.48            & 0.03   & 20-22700   & low   \\
J0837+0610   & \textbf{-6.01}   & 0.04   & 20-4850    & low   \\
J1840+5640   & -1.11            & 0.12   & 20-4850    & dou   \\
J0056+4756   & -0.89            & 0.14   & 20-4850    & high  \\
J1532+2745   & \textbf{-8.00}   & 6.88   & 25-1420    & dou   \\
J1635+2418   & \textbf{-5.40}   & 0.26   & 25-1420    & low   \\
J0934-5249   & -2.39            & 0.14   & 300-3100   & sim   \\
J1739-3023   & 0.00             & 0.03   & 300-3100   & high  \\
J1845-0743   & \textbf{-7.13}   & 0.04   & 300-3100   & low   \\
J1841-0425   & 0.00             & 0.09   & 300-3100   & high  \\
J1741-3016   & \textbf{-6.32}   & 0.06   & 300-3100   & dou   \\
J1941-2602   & \textbf{-4.54}   & 0.80   & 350-3100   & low   \\
J0818-3232   & -2.80            & 1.20   & 350-3100   & low   \\
J1831-0823   & -2.14            & 0.31   & 350-3100   & low   \\
J1722-3712   & -1.79            & 0.13   & 350-3100   & sim   \\
J1844-0433   & -1.78            & 0.09   & 350-3100   & sim   \\
J0820-3826   & -1.77            & 0.09   & 350-3100   & sim   \\
J2124-3358*  & -1.68            & 0.10   & 350-3100   & low   \\
J1741-3927   & -1.28            & 0.10   & 350-3100   & sim   \\
J1932-3655   & -1.23            & 0.20   & 350-3100   & sim   \\
J1904+0004   & -1.12            & 0.14   & 350-3100   & sim   \\
J1909-3744*  & -1.04            & 0.09   & 350-3100   & sim   \\
J1759-3107   & -0.94            & 0.19   & 350-3100   & high  \\
J1738-3211   & -0.88            & 0.18   & 350-3100   & sim   \\
J1700-3312   & -0.78            & 0.18   & 350-3100   & high  \\
J1708-3426   & -0.55            & 0.84   & 350-3100   & bro   \\
J1817-3618   & -0.55            & 0.16   & 350-3100   & high  \\
J1817-3837   & -0.54            & 0.15   & 350-3100   & high  \\
J0656-2228   & -0.36            & 0.24   & 350-3100   & sim   \\
J1901-0906   & -0.26            & 0.23   & 350-3100   & high  \\
J1727-2739   & 0.28             & 0.65   & 350-3100   & bro   \\
J1635-5954   & \textbf{-5.34}   & 0.70   & 400-3100   & low   \\
J1806-1154   & -3.70            & 1.28   & 400-3100   & low   \\
J1017-5621   & -2.64            & 0.27   & 400-3100   & low   \\
J1305-6455   & -2.35            & 0.38   & 400-3100   & sim   \\
J1157-6224   & -2.35            & 0.21   & 400-3100   & sim   \\
J1639-4604   & -2.34            & 0.26   & 400-3100   & sim   \\
J1326-6408   & -2.33            & 0.15   & 400-3100   & sim   \\
J1717-4054   & -2.28            & 0.43   & 400-3100   & sim   \\
J1046-5813   & -2.12            & 0.17   & 400-3100   & sim   \\
J1615-5537   & -2.08            & 0.30   & 400-3100   & sim   \\
J1123-6259   & -2.06            & 0.11   & 400-3100   & sim   \\
J1112-6613   & -2.00            & 0.14   & 400-3100   & sim   \\
J1711-5350   & -1.99            & 0.25   & 400-3100   & sim   \\
J1401-6357   & -1.96            & 0.19   & 400-3100   & sim   \\
J1600-5751   & -1.96            & 0.12   & 400-3100   & sim   \\
J1513-5908   & -1.93            & 0.19   & 400-3100   & low   \\
J1651-5222   & -1.90            & 0.29   & 400-3100   & sim   \\
J1312-5516   & -1.85            & 0.20   & 400-3100   & sim   \\
J1012-5857   & -1.76            & 0.09   & 400-3100   & sim   \\
J1810-5338   & -1.66            & 0.29   & 400-3100   & sim   \\
J0536-7543   & -1.66            & 0.21   & 400-3100   & sim   \\
J1016-5345   & -1.64            & 0.25   & 400-3100   & sim   \\
J1637-4553   & -1.45            & 0.13   & 400-3100   & sim   \\
J1646-6831   & -1.43            & 0.33   & 400-3100   & sim   \\
J1259-6741   & -1.35            & 0.22   & 400-3100   & sim   \\
J0711-6830*  & -1.33            & 0.07   & 400-3100   & sim   \\
J0745-5353   & -1.15            & 0.20   & 400-3100   & high  \\
J1326-6700   & -1.07            & 0.22   & 400-3100   & sim   \\
J1047-6709   & -0.62            & 0.22   & 400-3100   & sim   \\
J2129-5721*  & \textbf{-5.05}   & 3.09   & 436-3100   & low   \\
J1603-7202*  & -3.41            & 0.95   & 436-3100   & low   \\
J1210-5559   & -2.79            & 0.22   & 436-3100   & sim   \\
J1848-1414   & -2.34            & 0.30   & 436-3100   & sim   \\
J1549-4848   & -1.99            & 0.24   & 436-3100   & sim   \\
J1253-5820   & -1.53            & 0.29   & 436-3100   & sim   \\
J1822-4209   & -1.51            & 0.45   & 436-3100   & sim   \\
J0024-7204C* & \textbf{-8.00}   & 5.15   & 436-3968   & low   \\
J0024-7204D* & \textbf{-5.63}   & 0.08   & 436-3968   & dou   \\
J1844-0244   & \textbf{-8.00}   & 4.27   & 606-1606   & low   \\
J1826-1131   & -2.50            & 0.15   & 606-1606   & sim   \\
J1845-0434   & 0.00             & 0.62   & 606-1606   & high  \\
J1918+1444   & \textbf{-6.71}   & 2.42   & 606-4850   & low   \\
J1824-1118   & -2.20            & 0.11   & 606-4850   & sim   \\
J1820-1346   & -1.90            & 0.18   & 606-4850   & sim   \\
J1901+0716   & -1.81            & 0.13   & 606-4850   & sim   \\
J1808-2057   & -0.97            & 0.19   & 606-4850   & high  \\
J0218+4232*  & \textbf{-5.60}   & 0.11   & 74-1400    & low   \\
J1959+2048*  & -2.91            & 0.07   & 74-1400    & sim   \\
J1843-1113*  & -2.64            & 0.05   & 74-1400    & sim  \\

\end{longtable}
\begin{threeparttable}
\begin{tablenotes}
\small
\item[1] The asterisk (*) represents MSPs.
\item[2] Here, abbreviations are used for the names of the models, and the names after the abbreviations are indicated in parentheses: broken power law (bro), double turn-over spectrum (dou), and low-frequencies turn-over (low).
\end{tablenotes}
\end{threeparttable}

\subsection{Spectral index correlations}

The spectral index of pulsars serves as a characterization of how the flux density evolves with frequency, making it a prominent feature. Additionally, pulsars exhibit characteristic parameters that represent their unique properties. As a result, it is crucial to examine the correlation between spectral index and characteristic parameters of pulsars. The correlation between the spectral index, age, and period of $343$ pulsars was investigated by \cite{lorimer1995}. It was found that the spectral indices of millisecond pulsars exhibit a stronger correlation with the period $(r = +0.51)$ in comparison to normal pulsars. Although the correlation for characteristic age was weak for normal pulsars $(r = -0.19)$ and overall $(r = -0.22)$, they suggested that the characteristic age significantly impacts the spectral index more than the period: young pulsars have predominantly flatter spectra (typically -1) compared to older pulsars for which the corresponding value is 2.

\cite{Han2016} observed a strong correlation between the spectral index and the spin-down luminosity ($r = 0.32$) . The results of \cite{Han2016} were supported by \cite{jankowski2018}, who found that, among the $276$ pulsars they studied, the strongest correlation was between the spectral index and the spin-down luminosity ($r = 0.44$). Similarly, \cite{Zhao2019} calculated the Spearman correlation coefficients between characteristic parameters in a subset of pulsars ($19$ with simple power-law spectra, and $4$ with broken power-law spectra) and reported a strong correlation between the spectral index and the spin-down luminosity ($r = 0.62$). No correlation coefficient exceeded $0.5$ was found in their studies among other parameters.

Our research reveals no strong correlation between the spectral index and the characteristic parameters of both millisecond (see in Figure \ref{fig:3.2.5.1}) and normal pulsars (see in Figure \ref{fig:3.1.5.2}) when the model was not specified. However, when imposing restrictions and separately analyzing the correlation between the characteristic parameters of normal and millisecond pulsars under different models, significant diversity in the results emerged.

In comparison to normal pulsars (see Table \ref{tab:table4}), the correlation between the spectral index and specific characteristic parameters of millisecond pulsars is stronger (see Table \ref{tab:table3}). Specifically, in the case of a simple power-law spectrum, the pulsar distance (Dist) exhibits the strongest correlation with the spectral index ($r = -0.44$), while in the low-frequency turn-over model, the strength of the magnetic field at light cylinder ($B_{\rm {\text{LC}}}$) exhibits the strongest correlation with the spectral index ($r = -0.68$). This phenomenon is likely associated with the size of our sample. Despite having a relatively large sample size compared to previous studies, the number of detected pulsars remains limited. Thus, a strong correlation between the spectral index and certain characteristic parameters is anticipated. Consistent with our previous studies, both millisecond pulsars and normal pulsars exhibit a more pronounced correlation between the spectral index and certain characteristic parameters when the sample size is restricted. Nevertheless, additional sample data is required to validate this assertion.

\subsection{Deviations from a simple power-law spectrum}

Table \ref{tab:table5} presents our data set, covering a frequency range from $30$ MHz to $150$ GHz, that adequately captures the major frequency spectrum. To identify pulsars effectively, we require strict low-frequency and high-frequency coverage criteria. We consider pulsars as having a good low-frequency coverage when they have at least two data points below $600$ MHz. Similarly, pulsars with a good high-frequency coverage have at least one data point above $4$ GHz, as proposed by \cite{jankowski2018}.

\begin{table*}
\centering
\caption{\textbf{Fraction of pulsar spectra that can best be characterized by the
given spectral model. The first row provides total number of pulsars along with the total number and the second row gives the percentage (out of 886) for each model. The total number and percentage of pulsars with good low-frequency coverage (510) and high-frequency coverage (170) out of the total are presented in the next two columns. The remaining rows indicate the number and percentage (in the third column) of pulsars in the current model (first column) as a proportion of the total, as well as the percentage within the subset of pulsars with good low-frequency coverage (fifth column) and high-frequency coverage (seventh column).}}
\begin{threeparttable}
\label{tab:table5}
\begin{tabular}{lcccccc} % four columns, alignment for each
\hline
Set &pulsars&percent&\makecell[cc]{low-frequency\tnote{1}\\ coverage}&percent&\makecell[cc]{high-frequency\tnote{2}\\ coverage}&percent\\
\hline
Total&886&-&510&57.56&170&19.19\\
Simple power law & 607&68.51&295 & 57.84 & 53 & 31.18\\
Double-turn-over spectrum &23&2.60&20  & 3.92  & 13 & 7.65  \\
Broken power law & 54&6.09&50  & 9.80  & 31 & 18.24\\
low-frequency  turn-over&95&10.72&69  & 13.53 & 48 & 28.24\\
high-frequency  cut-off&107&12.08&76  & 14.90 & 25 & 14.71\\
\hline
\end{tabular}
\begin{tablenotes}
\item[1] at least two data points below 600 MHz for pulsars
\item[2] at least one data point above 4 GHz
\end{tablenotes}
\end{threeparttable}
\end{table*}

Our study achieved a good high-frequency coverage of 19.19$\%$ and a good low-frequency coverage of 57.56$\%$. We observed that some models' correlation coefficients changed significantly when a good low-frequency coverage was present, while the overall model ranking remaining the same. The Simple Power Law model correlation coefficient decreased from 68.51$\%$ to 57.56$\%$ (see Table \ref{tab:table5}), while the remaining four models had an increment in percentage between 1.5$\%$ to 3.8$\%$, indicating a shift in the correlation coefficients of spectral models. A good low-frequency coverage produced deviations from the simple power law spectral model.

With a good high-frequency coverage, we observed considerable changes that impacted the Simple Power Law model prominently. Its correlation coefficient decreased to 31.18$\%$, almost half of the original correlation coefficient. However, the low-frequency turn-over model's correlation coefficient increased to 28.24$\%$, almost doubling the original value, which almost equals the correlation coefficient of the top-ranked Simple Power Law model. The other three models exhibit variable increases in correlation coefficient, with the high-frequency cut-off Inversion model's minimal increase is 2.8$\%$.
We conclude that deviations from the simple power law spectral model are more pronounced with a good high-frequency coverage.

We observed significant changes in the spectral index with respect to the frequency coverage in Table \ref{tab:table6}. Firstly, we excluded the broken power law from our study, because it has two distinct spectral indices before and after the turn-over. In cases where there is adequate low frequency coverage ($-1.80\pm1.30$) and high frequency coverage ($-1.94\pm1.73$), the mean spectral index ($-1.77\pm1.31$) remains close to the standard spectral index ($-1.8$) \citep{maron2000}. If there is good coverage at both low and high frequencies, the spectral index becomes significantly steeper ($-2.01\pm1.16$). Conversely, when there is no good coverage of both low and high frequencies, the spectral index value is the highest ($-1.73\pm1.17$).

\textbf{When considering the impact of frequency coverage, we define a pulsar as having good low-frequency coverage when it has at least two data points below 600 MHz, and good high-frequency coverage when it has at least one data point exceeding 4 GHz. It was observed that the proportion of pulsars conforming to a simple power-law model was 57.84\% in the presence of good low frequency coverage, while this percentage dropped to 31.18\% in the case of good high frequency coverage. This suggests that pulsars are more inclined to diverge from the simple power-law model as they are observed at higher frequencies. Consequently, a focused analysis was performed on pulsars with frequencies ranging between 600 MHz and 4 GHz, which exhibit inadequate coverage in both low and high frequencies. Among the selected 313 pulsars, the proportion of single power-law models has reached 91.37\%. It is posited that with an expanded frequency coverage, the correspondence between the flux density and the frequency may deviate from a simple power-law model. Additionally, when studying the spectral index under different frequency coverage, the spectrum of the pulsar becomes steeper in a wider frequency range (below 600 MHz and above 4 GHz). Of course, there are only 92 objects in our data set that meet this condition, which is far short of a definitive conclusion. For both cases mentioned above, in the future, we need to observe more pulsars at lower and higher frequencies.}

\begin{table}
\caption{\textbf{The spectrum under different frequency coverage. Low-frequency coverage and high-frequency coverage are described in Table \ref{tab:table5}. Good-frequency coverage means simultaneously having good high-frequency and low-frequency coverage. None-frequency coverage means simultaneously lacking good high-frequency and low-frequency coverage. Total coverage means the spectral index in all models that do not include a broken power law.}}%
\centering%
\begin{threeparttable}
\setlength{\tabcolsep}{5mm}{
\begin{tabular}{lcccccc}%
\toprule%
Set &mean\\%\makecell[cc]{a\\ b}
\midrule%
low-frequency coverage&$-1.80\pm-1.30$\\
high-frequency coverage&$-1.94\pm-1.73$\\
good-frequency coverage&$-2.01\pm-1.56$\\
none-frequency coverage&$-1.73\pm -1.17$\\
total coverage&$-1.77\pm-1.31$\\
\bottomrule%
\end{tabular}}
\end{threeparttable}
\label{tab:table6}
\end{table}

\subsection{Implications of Spectral Turn-over and Indexes for Physical Models of Pulsar Magnetospheres}

The cutoffs at higher frequencies and turn-overs at lower frequencies seen in the spectra of pulsars are inherent characteristic of the physical conditions prevailing inside their magnetospheres. There is considerable consensus that the radio emission of pulsars originates from radiation of the charged particle bunches accelerated along the curved magnetic field lines \citep{Sturrock1971,Ruderman1975,Philippov2022}. The pulsar magnetospheres are thought to be a tenuous weakly turbulent electron-positron plasma within which various nonlinear processes associated with instabilities and resonances operate under strong magnetic fields to produce electromagnetic waves. \citet{Machabeli2005} have shown that scattering of the longitudinal Langmiur plasma waves by the drift of charged particles in the electric field generates a spectrum with index $-1.5$, which is in qualitative agreement with what found from fitting to the pulsar spectra with a simple power-law: $\alpha=-1.57\pm0.32$ for normal pulsars and $\alpha=-1.89\pm0.21$ for millisecond pulsars. Refraction of the electromagnetic waves is another key ingredient modifying and re-shaping the formed pulsar spectrum as it is frequency depended \citep{Petrova2002}. The kinetic energy of the magnetospheric plasma particles acquired during the process of acceleration along the curved magnetic field lines is comparable to the observed brightness temperatures of pulsars and therefore self-absorption of the emitted curvature radiation severely limits the formed spectra. \citet{Usov1984} suggested that both the breaks and the low frequency turn-overs in the spectra observed from the pulsars arise from the self-absorption of the curvature radiation by emitting bunches themselves. This mechanism naturally accounts for the exponential-like shape of the turn-over. The reason of the appearance of two characteristic frequencies in the spectrum is due to the non-uniformity and anisotropy of the radiation source. These break ($\nu_{\rm b}$) and turn-over ($\nu_{\rm peak}$) frequencies are given by
\begin{equation}
    \nu_{\rm b}=\left[\frac{3}{4}\left(\frac{2}{3}\right)^{2/3}\frac{(k-2)e^{2}\theta^{2(k-2)}_{\rm p}\dot{N}}{m_{\rm e}cR^{2}\Gamma}\left(\frac{8}{9}\frac{R}{c\Gamma^{3}}\right)^{-k}\right]^{1/(2+k)},
\end{equation}
and 
\begin{equation}
    \nu_{\rm peak}=\nu_{\rm b}\theta^{[16/3(2+k)]}_{\rm p},
\end{equation}
respectively. Here, $k$ is a constant, $e$ and $m_{\rm e}$ are the charge and the mass of electron, $\dot{N}$ is the number of particles ejected from the pulsar per unit time, $c$ is the speed of light, $\Gamma$ is the Lorentz factor, $R$ is the radius of the neutron star and $\theta_{\rm p}=(R\Omega/c)$ is the angular radius of the polar cap with $\Omega=2\pi/P$ being the pulsar's angular rotation velocity. Their model predicts a spectrum with $\propto\nu^{-4(k-2)/(6+k)}$ before the break and with a much steeper slope $\propto\nu^{-(k-2)}$ after the break. The results presented in Sections \ref{subsec:bplnp} and \ref{subsec:bplmp} agree well with $k=4$ and $k=5$, respectively (see table 1 in \citet{Usov1984}). The flattening of the pulsar spectra can be understood in terms of the geometry of the hollow-cone model \citep{Kuzmin1986}, as shrinking of the hollow cone width with increasing frequency causes steepening of the emergent radiation. According to \citet{Kontorovich2013}, the high frequency cutoff of the pulsar spectra is a consequence of acceleration of the electrons to the speed of light in the huge electrostatic potential present above the polar cap. In contrast to the conventional models \citep{Sturrock1971,Ruderman1975}, their theory assumes a zero potential at the neutron star surface, a linear increasing electrostatic field just above the polar cap which passes through a maximum slowly and then experiences rapid decline when electrons approach to the relativistic limit. This results in a flat spectrum radiated by a large number of electrons accelerated in the corresponding potential across the polar gap. That theoretical model predicts a cutoff frequency
\begin{equation}
    \nu_{\rm c}\approx \sqrt{2\left(\frac{B}{2\times10^{12}\,\mbox{G}}\right)\left(\frac{1\,\mbox{s}}{P}\right)}\,\mbox{GHz},
    \label{kf2013}
\end{equation}
while empirical relation found earlier by \citet{Malov1981} states that the cutoff position in the spectrum is
\begin{equation}
    \nu_{\rm c}\approx1.4\left(\frac{1\,\mbox{s}}{P}\right)^{0.46\pm0.18}\,\mbox{GHz}.
\end{equation}
Also their model is capable of explaining a low frequency turn-over occurring naturally at $\nu_{\rm peak}\approx0.1\nu_{\rm c}$ \citep{Kontorovich2013b}. Equation (\ref{kf2013}) may, in principle, account for the rarity of pulsars showing cutoff behaviour in their spectra (see Table \ref{tab:table00}), since for most of them the emission region shifts to a thin and narrow layer.   
The absence of a low frequency turn-over in the spectrum, especially in the case of millisecond pulsars, can be attributed to deviation of the magnetic field configuration from a dipolar structure or to compactness of the emission region \citep{Kuzmin2001}. In addition to these, other geometrical factors like curvature radius of the magnetospheric field lines, altitude of the emission region above the polar cap and orientation of the magnetic dipolar axis with respect to the rotational axis affect the spectral shape as well and may lead to high energy cutoff observed from a given pulsar \citep{Malov1991}. Another possibility applicable for spectral turn-over at high frequencies seen from GPS pulsars is the free-free absorption of the pulsar emission by the interstellar matter with HII regions, pulsar wind nebulae and supernova remnants being the primary sources \citep{kijak2017}. For example, PSR J1056--6258 is known to be surrounded by an ionized hydrogen region \citep{Koribalski1995}. Another case supporting the view that environment may cause inversion of the spectra of pulsars at high frequencies is PSR B1259--63 which orbits around a Be star. It is observed that the pulsar's spectrum evolves regularly as it approaches to the companion star during binary motion \citep{Kijak2011a}. Model fitting to the temporal evolution of the pulsar spectra with the free-free absorption may in principle enable one to estimate the electron density and the size of the absorbing medium. For instance, by using this method \citet{Basu2016} constrained the physical properties of the medium enclosures PSR B1800--21. By examining 33 GPS pulsars, \citet{Li2023} arrived at the conclusion that their spectral indices obey a bimodal distribution. They also found a strong positive relation between periods $P$ and DMs of GPS pulsars. This suggests that at least some of the GPS pulsars may be evolved through interaction with a fallback disk for which there exists interplay between the slow-down torque and the amount of the dispersive matter settled down around the neutron star after the supernova explosion \citep{Alpar2001,Menou2001}.

\section{Conclusions}
\label{sec:conclusions}

In our work, we combined an open-source database developed by \cite{swainston2022} to determine flux density values for a total of 941 pulsars. The total covered frequency range was from $30$ MHz to $150$ GHz, and each pulsar had at least $4$ different flux density values. After excluding some pulsars that cannot be fitted, we determined the spectral values of $886$ pulsars. Subsequently, we applied five distinct spectral models to fit the data. Our results can be summarized as follows:
\begin{itemize}
    \item \textbf{The spectral indexes of our sample is normally distributed, with a weighted mean is $-1.61 \pm 0.32$.}
    \item Overall, $68.51$\% of the pulsars fit within the range of the simple power law model, while $31.5$\% deviate from it, showing other characteristics. Among them, $12.08$\% showed high-frequency cut-off, and $10.72$\% showed low-frequency turn-over characteristics.
    \item At both high and low frequencies, pulsars tend to deviate more from a simple power-law spectrum and have a steeper spectral index.
    \item Compared to normal pulsars, millisecond pulsars exhibit stronger correlations with characteristic parameters. \textbf{It is noteworthy to keep in mind that this research examined only a total of 86 millisecond pulsars.}
    \item In our statistical work, 33 GPS pulsars have been identified. Of these, 11 had been mentioned in previous literature, and 22 were identified for the first time.
\end{itemize}

%% Also note that the akcnowlodgment environment does not support long amounts of text. If you have a lot of people and institutions to acknowledge, do not use this command. Instead, create a new \section{Acknowledgments}.
\begin{acknowledgments}
We gratefully acknowledge the financial support received from various funding sources for this work. This includes the National Natural Science Foundation of China (Grants No. 12103013, 11988101, U1731238, U2031117, 11565010, 11725313, 1227308, 12041303), the Foundation of Science and Technology of Guizhou Province (Grants No. (2021)023, (2016)4008, (2017)5726-37), the Foundation of Guizhou Provincial Education Department (Grants No. KY(2020)003, KY(2021)303, KY(2023)059), the National SKA Program of China (Grants No. 2022SKA0130100, 2022SKA0130104, 2020SKA0120200), the Youth Innovation Promotion Association CAS (id. 2021055), CAS Project for Young Scientists in Basic Research (grant YSBR-006), Foreign Talents Programme with the Grant QN2023061004L (EG), the CAS Youth Interdisciplinary Team ,the CAS Youth Interdisciplinary Team and the Cultivation Project for FAST Scientific Payoff and Research Achievement of CAMS-CAS. Their support has been instrumental in the successful completion of this work.

\end{acknowledgments}

\renewcommand{\thetable}{A\arabic{table}}
\setcounter{table}{0}
\begin{longtable}{@{}p{2cm}cccccccp{8cm}@{}}
    \caption{\textbf{Pulsars that have broken power-law spectra, where $\nu_{br}$ is the frequency of the spectral break, $\alpha_1$ and $\alpha_2$ the spectral index before and after the break for a given frequency range. They are followed by corresponding errors. We also mark the MSPs with*.}\label{tab:a1}} \\
    \hline
    PSRJ       & $\alpha_1$& $\alpha_1$\_err & $\alpha_2$& $\alpha_2$\_err & $\nu_{br}$(MHz)& $\nu_{br}$\_err(MHz) & Freq.range(MHz) \\
    \hline
    \endfirsthead

\multicolumn{8}{c}%
{{\bfseries \tablename\ \thetable{} continue}} \\
\hline
 PSRJ       & $\alpha_1$ & $\alpha_{1\text{ err}}$ & $\alpha_2$ & $\alpha_{2\text{ err}}$ & $\nu_{\text{br}}$ (MHz) & $\nu_{\text{br err}}$ (MHz) & Freq. range (MHz) \\
\hline
\endhead

\hline
\multicolumn{8}{r}{{Continued on next page}} \\
\endfoot

\hline
\endlastfoot

J0040+5716	&	-0.96 	&	0.51 	&	-1.74 	&	0.08 	&	408	&	0.2	&	102.5-4850	\\
J0147+5922	&	-1.52 	&	0.13 	&	-0.52 	&	0.13 	&	760.9	&	189.4	&	102.5-4920	\\
J0437-4715*	&	-0.92 	&	0.01 	&	-1.71 	&	0.15 	&	1400.1	&	489.3	&	76-17000	\\
J0452-1759	&	-0.34 	&	0.07 	&	-2.13 	&	0.11 	&	699.9	&	64.5	&	102.5-4850	\\
J0543+2329	&	-0.21 	&	0.15 	&	-1.53 	&	0.12 	&	814.6	&	90	&	102.5-10550	\\
J0612+3721	&	-0.03 	&	0.45 	&	-2.05 	&	0.17 	&	447.5	&	70.9	&	59.66-4850	\\
J0742-2822	&	-1.15 	&	0.09 	&	-1.57 	&	0.12 	&	1582.2	&	669.1	&	102.5-10550	\\
J0820-4114	&	-0.67 	&	0.23 	&	-2.32 	&	0.42 	&	500.7	&	151	&	76-3100	\\
J1001-5507	&	-0.51 	&	0.52 	&	-2.58 	&	0.28 	&	567.6	&	190.3	&	151.5-3100	\\
J1024-0719*	&	-1.42 	&	0.09 	&	-0.64 	&	0.19 	&	1400	&	0.6	&	102-4850	\\
J1059-5742	&	-1.06 	&	0.44 	&	-2.64 	&	0.35 	&	684.1	&	264.2	&	154.24-3100	\\
J1141-6545	&	3.00 	&	2.14 	&	-2.57 	&	0.26 	&	254.5	&	18	&	154.24-3100	\\
J1146-6030	&	-0.37 	&	0.23 	&	-1.62 	&	0.24 	&	600	&	22.4	&	154.24-3100	\\
J1224-6407	&	-0.80 	&	0.10 	&	-1.54 	&	0.15 	&	843	&	0.5	&	154.24-3100	\\
J1300+1240	&	3.00 	&	7.17 	&	-2.53 	&	0.14 	&	105.7	&	6.7	&	50-1400	\\
J1326-5859	&	0.94 	&	2.31 	&	-1.88 	&	0.25 	&	647.1	&	201.7	&	400-8356	\\
J1453+1902*	&	-0.98 	&	0.22 	&	-2.68 	&	1.32 	&	480	&	346.9	&	149-2100	\\
J1455-3330*	&	-1.57 	&	0.36 	&	-2.83 	&	0.39 	&	660	&	3	&	154.24-1520	\\
J1518+4904	&	-0.19 	&	0.66 	&	-2.44 	&	0.32 	&	1156.6	&	297.3	&	102-4850	\\
J1600-5044	&	-0.63 	&	0.08 	&	-2.05 	&	0.08 	&	728	&	0.2	&	147.5-8356	\\
J1604-4909	&	-0.93 	&	0.16 	&	-2.47 	&	0.42 	&	950	&	0.3	&	147.5-3100	\\
J1633-5015	&	-1.04 	&	0.10 	&	-2.51 	&	0.24 	&	843	&	0.4	&	147.5-3100	\\
J1643-1224*	&	-1.79 	&	0.04 	&	-1.38 	&	0.09 	&	1284	&	0.2	&	147.5-4850	\\
J1644-4559	&	2.50 	&	0.10 	&	-2.07 	&	0.00 	&	729.9	&	3	&	147.5-17000	\\
J1651-4246	&	-2.16 	&	0.01 	&	-2.47 	&	0.12 	&	1315.9	&	58.3	&	147.5-5124	\\
J1705-1906	&	-1.11 	&	0.11 	&	-0.30 	&	0.32 	&	4850	&	17.1	&	102.5-8600	\\
J1708-3426	&	-0.55 	&	0.84 	&	-2.99 	&	0.31 	&	816	&	130.4	&	350-3100	\\
J1713+0747*	&	-0.13 	&	0.07 	&	-2.55 	&	0.16 	&	2030	&	81	&	102-4850	\\
J1727-2739	&	0.28 	&	0.65 	&	-2.07 	&	0.19 	&	939.9	&	231.6	&	350-3100	\\
J1739-3131	&	-0.61 	&	0.06 	&	-4.78 	&	0.81 	&	1369	&	0.1	&	300-1606	\\
J1803-2137	&	1.70 	&	0.11 	&	-0.59 	&	0.03 	&	835.2	&	33.5	&	300-8600	\\
J1807-0847	&	-0.03 	&	0.27 	&	-1.62 	&	0.12 	&	843	&	1.4	&	102.5-8600	\\
J1809-1917	&	0.96 	&	0.26 	&	-1.45 	&	0.37 	&	1908.2	&	231.1	&	325-6591	\\
J1813+4013	&	-0.97 	&	0.16 	&	-2.91 	&	0.60 	&	862.5	&	228.8	&	59.67-2600	\\
J1816+4510*	&	0.94 	&	2.03 	&	-2.54 	&	0.38 	&	142.5	&	0.4	&	74-820	\\
J1823-1115	&	0.21 	&	0.19 	&	-1.56 	&	0.13 	&	628.4	&	56.1	&	300-4850	\\
J1833-0338	&	-0.98 	&	0.10 	&	-3.01 	&	0.13 	&	475.5	&	24.3	&	102.5-3100	\\
J1841+0912	&	-0.97 	&	1.13 	&	-1.95 	&	0.09 	&	296.9	&	388.7	&	59.73-4850	\\
J1844+1454	&	-1.27 	&	0.80 	&	-4.40 	&	0.49 	&	102.5	&	0.6	&	49.8-1420	\\
J1849-0636	&	2.23 	&	2.18 	&	-2.32 	&	0.09 	&	157.9	&	7.5	&	102.5-3100	\\
J1857+0943*	&	-1.50 	&	0.07 	&	-0.01 	&	0.11 	&	1405.1	&	12.6	&	102-4850	\\
J1907+4002	&	-0.32 	&	0.24 	&	-2.07 	&	0.12 	&	480.4	&	72.3	&	100-4850	\\
J1909+1102	&	1.07 	&	1.94 	&	-2.54 	&	0.10 	&	178.7	&	20.6	&	102.5-4920	\\
J1918-0642*	&	-1.59 	&	0.06 	&	-8.00 	&	0.40 	&	1284	&	0.2	&	149-1400	\\
J1935+1616	&	-0.45 	&	0.06 	&	-3.32 	&	0.66 	&	2291.9	&	648.1	&	135.25-22700	\\
J1939+2134	&	-1.51 	&	0.02 	&	-2.93 	&	0.21 	&	1805.6	&	227.8	&	74-4850	\\
J2022+2854	&	1.03 	&	0.61 	&	-1.29 	&	0.10 	&	100.2	&	21.6	&	24.75-22700	\\
J2055+3630	&	-0.65 	&	0.35 	&	-1.97 	&	0.08 	&	454.6	&	30.7	&	102.5-4850	\\
J2149+6329	&	0.06 	&	0.19 	&	-2.10 	&	0.13 	&	472.3	&	23.8	&	102.5-4850	\\
J2155-3118	&	-1.27 	&	0.23 	&	-2.60 	&	0.22 	&	534.7	&	129.5	&	147.5-1408	\\
J2229+6205	&	-0.28 	&	0.79 	&	-2.30 	&	0.16 	&	235.6	&	59.8	&	102.5-1410	\\
J2308+5547	&	-0.81 	&	0.72 	&	-8.00 	&	2.48 	&	236.1	&	25.8	&	59.66-1420	\\
J2317+2149	&	0.47 	&	0.38 	&	-2.05 	&	0.09 	&	198.1	&	21.4	&	25-4850	\\
J2321+6024	&	0.05 	&	0.26 	&	-2.97 	&	0.41 	&	1161.7	&	66.9	&	102.5-22700	\\

\end{longtable}

\begin{longtable}{@{}p{2cm}cccccccp{8cm}@{}}
    \caption{\textbf{\textbf{Pulsars that have power-law spectra with a low-frequency turn-over, where $\nu_p$ is the peak/turn-over frequency, $\alpha$ is the spectral index for a given frequency range, and $\beta$ is the free fit parameter that determines the smoothness of the turn-over. They are followed by corresponding errors. We also mark the MSPs with*. }}\label{tab:a3}} \\
    \hline
   PSRJ        & $\alpha$& $\alpha$\_err & $\beta$& $\beta$\_err & $\nu_p$(MHz)  & $\nu_p$\_err(MHz) & Freq.range(MHz)\\
    \hline
    \endfirsthead

\multicolumn{8}{c}%
{{\bfseries \tablename\ \thetable{} continue}} \\
\hline
PSRJ        & $\alpha$& $\alpha$\_err & $\beta$& $\beta$\_err & $\nu_p$(MHz)  & $\nu_p$\_err(MHz) & Freq.range(MHz)\\
\hline
\endhead

\hline
\multicolumn{8}{r}{{Continued on next page}} \\
\endfoot

\hline
\endlastfoot

J0024-7204C*	&	-8.00 	&	5.15 	&	0.38 	&	0.00 	&	600.4	&	9	&	436-3968	\\
J0024-7204J*	&	-8.00 	&	0.34 	&	2.10 	&	0.08 	&	607.8	&	4.5	&	768-3968	\\
J0151-0635	&	-0.80 	&	0.18 	&	2.10 	&	1.94 	&	71	&	10.6	&	25-1408	\\
J0218+4232*	&	-5.60 	&	0.11 	&	2.10 	&	0.33 	&	174.2	&	0.7	&	74-1400	\\
J0304+1932	&	-7.93 	&	0.10 	&	0.12 	&	0.00 	&	184.1	&	11.6	&	24.75-4850	\\
J0323+3944	&	-3.25 	&	1.36 	&	0.63 	&	0.44 	&	80.8	&	15.5	&	25-1408	\\
J0406+6138	&	-5.26 	&	0.07 	&	0.12 	&	0.00 	&	61.1	&	5.7	&	102.5-4850	\\
J0601-0527	&	-1.77 	&	0.10 	&	1.87 	&	1.83 	&	139.2	&	47.4	&	102.5-4850	\\
J0613-0200*	&	-2.46 	&	0.21 	&	1.41 	&	0.41 	&	466.6	&	46.3	&	102-3100	\\
J0614+2229	&	-2.20 	&	0.09 	&	2.10 	&	0.30 	&	216.7	&	9.4	&	85-4850	\\
J0621+1002*	&	-6.61 	&	0.12 	&	0.14 	&	0.00 	&	114.8	&	7.5	&	102.5-4850	\\
J0700+6418	&	-2.60 	&	0.41 	&	1.49 	&	0.49 	&	78.1	&	5.8	&	49.8-1408	\\
J0738-4042	&	-2.76 	&	0.02 	&	0.60 	&	0.00 	&	397.1	&	3.3	&	107-17000	\\
J0809-4753	&	-2.62 	&	0.23 	&	1.46 	&	0.40 	&	127.2	&	6.8	&	76-3100	\\
J0814+7429	&	-1.81 	&	0.05 	&	2.01 	&	1.42 	&	46.8	&	0.6	&	20-22700	\\
J0818-3232	&	-2.80 	&	1.20 	&	1.59 	&	1.69 	&	525	&	38.6	&	350-3100	\\
J0820-1350	&	-3.62 	&	0.66 	&	0.29 	&	0.05 	&	57.8	&	13.8	&	58.6-4850	\\
J0826+2637	&	-1.48 	&	0.03 	&	2.10 	&	0.24 	&	45.7	&	3.2	&	20-22700	\\
J0828-3417	&	-5.06 	&	0.53 	&	1.22 	&	0.27 	&	147.7	&	2.5	&	76-1400	\\
J0835-4510	&	-2.06 	&	0.18 	&	0.53 	&	0.14 	&	99.2	&	10.1	&	76-343500	\\
J0837+0610	&	-6.01 	&	0.04 	&	0.23 	&	0.00 	&	60.7	&	3.8	&	20-4850	\\
J0837-4135	&	-2.43 	&	0.20 	&	1.07 	&	0.21 	&	272.4	&	11.2	&	147.5-8600	\\
J0840-5332	&	-3.02 	&	2.18 	&	0.76 	&	1.26 	&	153.1	&	112.1	&	151.5-3100	\\
J0846-3533	&	-2.32 	&	0.24 	&	2.10 	&	1.51 	&	498.1	&	28.6	&	147.5-3100	\\
J0855-3331	&	-2.61 	&	0.67 	&	0.68 	&	0.52 	&	43.5	&	4.8	&	147.5-3100	\\
J0907-5157	&	-1.21 	&	0.11 	&	2.10 	&	1.99 	&	156.8	&	19	&	147.5-3100	\\
J0908-1739	&	-1.45 	&	0.07 	&	2.10 	&	1.33 	&	84.2	&	1.4	&	49.8-4850	\\
J0908-4913	&	-1.98 	&	0.41 	&	1.22 	&	0.77 	&	603.5	&	42.3	&	400-17000	\\
J0942-5552	&	-4.68 	&	0.03 	&	0.47 	&	0.00 	&	305.1	&	6.5	&	151.5-3100	\\
J0943+1631	&	-1.80 	&	0.56 	&	1.22 	&	1.47 	&	71.1	&	22.5	&	25-3100	\\
J1017-5621	&	-2.64 	&	0.27 	&	2.10 	&	2.00 	&	434.9	&	108.5	&	400-3100	\\
J1056-6258	&	-1.47 	&	0.23 	&	2.10 	&	1.38 	&	580.9	&	96	&	400-8356	\\
J1057-5226	&	-2.53 	&	0.46 	&	2.10 	&	1.66 	&	186.4	&	27.2	&	151.5-3100	\\
J1243-6423	&	-5.71 	&	0.57 	&	2.10 	&	2.00 	&	562.6	&	44.5	&	408-8356	\\
J1327-6222	&	-2.17 	&	0.26 	&	1.71 	&	1.11 	&	216.2	&	24.5	&	151.5-6591	\\
J1359-6038	&	-2.31 	&	0.14 	&	1.88 	&	1.88 	&	179.6	&	36.4	&	151.5-8356	\\
J1430-6623	&	-1.92 	&	0.20 	&	2.10 	&	1.30 	&	238.7	&	13.8	&	151.5-8356	\\
J1453-6413	&	-2.64 	&	0.17 	&	2.10 	&	0.15 	&	192.9	&	8.8	&	102-8356	\\
J1513-5908	&	-1.93 	&	0.19 	&	2.10 	&	1.37 	&	611.6	&	65.6	&	400-3100	\\
J1524-5706	&	-2.43 	&	0.46 	&	2.10 	&	1.43 	&	1189.4	&	65.2	&	728-3100	\\
J1559-4438	&	-5.16 	&	0.01 	&	0.34 	&	0.00 	&	403.8	&	2.7	&	147.5-5124	\\
J1600-3053*	&	-7.57 	&	0.05 	&	0.22 	&	0.00 	&	847	&	18.7	&	350-4820	\\
J1603-7202*	&	-3.41 	&	0.95 	&	1.65 	&	1.49 	&	594.2	&	28	&	436-3100	\\
J1614+0737	&	-2.09 	&	0.29 	&	1.38 	&	1.50 	&	52.3	&	7.8	&	25-4850	\\
J1623-2631	&	-7.97 	&	0.20 	&	0.31 	&	0.01 	&	395.2	&	21.8	&	106.5-2695	\\
J1635+2418	&	-5.40 	&	0.26 	&	0.30 	&	0.02 	&	73.2	&	8.5	&	25-1420	\\
J1635-5954	&	-5.34 	&	0.70 	&	2.10 	&	1.84 	&	857.2	&	59.9	&	400-3100	\\
J1651-5255	&	-8.00 	&	0.04 	&	0.40 	&	0.00 	&	473.2	&	15.8	&	40-3100	\\
J1705-3950	&	-1.13 	&	0.46 	&	2.10 	&	1.73 	&	1039.4	&	71.9	&	610-3100	\\
J1709-4429	&	-1.43 	&	0.09 	&	2.10 	&	1.41 	&	485.4	&	55	&	400-8600	\\
J1717-3425	&	-2.43 	&	0.24 	&	2.10 	&	1.57 	&	175.6	&	20.7	&	147.5-3100	\\
J1722-3207	&	-2.50 	&	0.36 	&	0.78 	&	0.32 	&	44.8	&	2.5	&	147.5-3100	\\
J1731-4744	&	-2.09 	&	0.08 	&	2.10 	&	0.29 	&	214.7	&	2.5	&	133-3100	\\
J1740-3015	&	-1.14 	&	0.17 	&	2.10 	&	0.32 	&	951.4	&	146.7	&	350-17000	\\
J1743-3150	&	-3.00 	&	0.20 	&	2.10 	&	1.27 	&	476.4	&	10.4	&	147.5-3100	\\
J1745-3040	&	-1.37 	&	0.09 	&	2.10 	&	0.21 	&	1009.9	&	130.2	&	350-10550	\\
J1751-3323	&	-1.40 	&	0.99 	&	1.76 	&	1.72 	&	1006.2	&	109.3	&	610-3100	\\
J1751-4657	&	-2.78 	&	0.18 	&	2.10 	&	1.45 	&	184.5	&	14.9	&	147.5-3100	\\
J1753-2501	&	-2.84 	&	2.04 	&	0.51 	&	0.45 	&	760.8	&	91.1	&	300-4850	\\
J1757-2421	&	-1.81 	&	0.14 	&	2.10 	&	0.17 	&	557.8	&	10.4	&	300-6591	\\
J1801-2304	&	-6.95 	&	0.12 	&	0.15 	&	0.00 	&	261.6	&	17.3	&	300-8000	\\
J1801-2920	&	-2.15 	&	0.45 	&	1.61 	&	1.75 	&	157.9	&	46.8	&	147.5-3100	\\
J1806-1154	&	-3.70 	&	1.28 	&	1.98 	&	1.94 	&	642.3	&	64.7	&	400-3100	\\
J1818-1422	&	-3.13 	&	0.19 	&	1.26 	&	0.08 	&	886.1	&	54	&	606-5124	\\
J1820-0427	&	-2.38 	&	0.04 	&	2.10 	&	0.05 	&	127.2	&	0.3	&	102.5-4920	\\
J1822-2256	&	-3.28 	&	0.24 	&	0.62 	&	0.07 	&	231.3	&	38.4	&	350-4850	\\
J1826-1334	&	-0.78 	&	0.52 	&	1.06 	&	0.57 	&	1220.1	&	134.6	&	325-4920	\\
J1830-1059	&	-8.00 	&	7.98 	&	0.36 	&	0.05 	&	913.1	&	47.8	&	606-4920	\\
J1831-0823	&	-2.14 	&	0.31 	&	2.10 	&	1.07 	&	450.5	&	40.8	&	350-3100	\\
J1832-0827	&	-3.55 	&	0.91 	&	0.84 	&	0.34 	&	573.3	&	27.5	&	300-6591	\\
J1833-0827	&	-4.35 	&	0.99 	&	0.20 	&	0.06 	&	394.8	&	126.6	&	300-4920	\\
J1835-1020	&	-2.78 	&	0.12 	&	0.65 	&	0.02 	&	587.7	&	31.1	&	325-4850	\\
J1836-1008	&	-6.69 	&	0.05 	&	0.26 	&	0.00 	&	209.6	&	6.6	&	147.5-3100	\\
J1843-0000	&	-1.83 	&	0.06 	&	2.10 	&	1.58 	&	472.3	&	21.1	&	350-5124	\\
J1843-0459	&	-3.78 	&	0.52 	&	2.10 	&	1.77 	&	807.4	&	59.1	&	728-3100	\\
J1844-0244	&	-8.00 	&	4.27 	&	0.89 	&	0.35 	&	861.1	&	45.7	&	606-1606	\\
J1844-0538	&	-2.10 	&	0.20 	&	2.10 	&	0.42 	&	1008	&	57.2	&	500-4850	\\
J1845-0743	&	-7.13 	&	0.04 	&	0.31 	&	0.00 	&	804.5	&	9.1	&	300-3100	\\
J1852-0635	&	-0.99 	&	0.03 	&	2.10 	&	0.10 	&	1119	&	41.1	&	610-8350	\\
J1905-0056	&	-4.31 	&	0.09 	&	0.29 	&	0.01 	&	74.4	&	7.4	&	102.5-3100	\\
J1909+0254	&	-2.78 	&	0.27 	&	1.93 	&	1.91 	&	136.6	&	33.1	&	102.5-1410	\\
J1913-0440	&	-2.24 	&	0.22 	&	1.09 	&	0.53 	&	48.3	&	4.3	&	102.5-4850	\\
J1918+1444	&	-6.71 	&	2.42 	&	2.09 	&	1.99 	&	1998.4	&	1640.8	&	606-4850	\\
J1941-2602	&	-4.54 	&	0.80 	&	2.10 	&	0.39 	&	484.6	&	14.6	&	350-3100	\\
J1943-1237	&	-2.22 	&	0.16 	&	2.10 	&	1.77 	&	159.6	&	17.7	&	147.5-1408	\\
J2002+4050	&	-8.00 	&	1.34 	&	0.20 	&	0.00 	&	247.2	&	9.5	&	102.5-4850	\\
J2004+3137	&	-2.70 	&	0.21 	&	2.10 	&	1.42 	&	424.4	&	26.3	&	350-4850	\\
J2046-0421	&	-2.73 	&	1.00 	&	0.99 	&	0.73 	&	193.6	&	21.7	&	102.5-3100	\\
J2113+2754	&	-1.89 	&	0.10 	&	2.10 	&	1.53 	&	42.6	&	4.9	&	25-4850	\\
J2124-3358*	&	-1.68 	&	0.10 	&	2.10 	&	1.28 	&	440.1	&	48	&	350-3100	\\
J2129-5721*	&	-5.05 	&	3.09 	&	1.22 	&	1.55 	&	567.6	&	65.6	&	436-3100	\\
J2219+4754	&	-3.23 	&	0.09 	&	0.50 	&	0.02 	&	41.6	&	4	&	35.1-1435	\\
J2305+3100	&	-1.44 	&	0.06 	&	2.10 	&	1.33 	&	94.5	&	7.7	&	49.8-4850	\\
J2313+4253	&	-1.39 	&	0.12 	&	1.73 	&	1.70 	&	58.9	&	15.4	&	25-10550	\\
J2330-2005	&	-2.05 	&	0.07 	&	2.10 	&	0.44 	&	75.7	&	0.8	&	35.1-4920	\\

\end{longtable}

\begin{longtable}{@{}p{4cm}cccccp{8cm}@{}}
    \caption{\textbf{\textbf{Pulsars that have power-law spectra with a high-frequency cut-off, where $\nu_{\rm c}$ is the cut-off frequency and $\alpha$ is the spectral index for a given frequency range. They are followed by corresponding errors. We also mark the MSPs with*. }}\label{tab:a4}} \\
    \hline
PSRJ        & $\alpha$& $\alpha$\_err & $\nu_{\rm c}$(MHz)  & $\nu_{\rm c}$\_err(MHz) & Freq.range(MHz)\\
    \hline
    \endfirsthead

\multicolumn{6}{c}%
{{\bfseries \tablename\ \thetable{} continue}} \\
\hline
PSRJ        & $\alpha$& $\alpha$\_err & $\nu_{\rm c}$(MHz)  & $\nu_{\rm c}$\_err(MHz) & Freq.range(MHz)\\
\hline
\endhead

\hline
\multicolumn{6}{r}{{Continued on next page}} \\
\endfoot

\hline
\endlastfoot

J0056+4756	&	-0.89 	&	0.14 	&	5984.60 	&	744.80 	&	20-4850	\\
J0102+6537	&	0.00 	&	0.49 	&	1944.60 	&	167.00 	&	102.5-1606	\\
J0108+6608	&	-1.16 	&	0.22 	&	2277.40 	&	725.00 	&	65-1408	\\
J0133-6957	&	-0.18 	&	0.17 	&	1431.50 	&	5.90 	&	154.24-1400	\\
J0357+5236	&	-0.60 	&	0.19 	&	2194.30 	&	346.30 	&	102.5-1606	\\
J0450-1248	&	-1.46 	&	0.12 	&	2376.80 	&	491.10 	&	102.5-1420	\\
J0502+4654	&	-0.35 	&	0.23 	&	1927.30 	&	193.10 	&	102.5-1420	\\
J0525+1115	&	-0.88 	&	0.11 	&	1884.90 	&	109.50 	&	61-1420	\\
J0533+0402	&	0.00 	&	4.93 	&	1525.80 	&	47.50 	&	102.5-1360	\\
J0737-3039A	&	-1.57 	&	0.06 	&	8136.30 	&	1698.80 	&	99-5000	\\
J0745-5353	&	-1.15 	&	0.20 	&	5765.20 	&	1538.80 	&	400-3100	\\
J0838-2621	&	-0.22 	&	0.34 	&	1689.00 	&	114.20 	&	350-1460.04	\\
J0904-7459	&	-0.83 	&	0.19 	&	4026.00 	&	876.20 	&	154.24-3100	\\
J0921+6254	&	-0.44 	&	0.21 	&	1731.00 	&	296.10 	&	24.75-1408	\\
J0955-5304	&	-0.69 	&	0.22 	&	1546.00 	&	40.70 	&	154.24-1440	\\
J1003-4747	&	-1.20 	&	0.10 	&	5211.10 	&	1205.10 	&	154.24-3100	\\
J1012+5307*	&	-0.85 	&	0.07 	&	5332.70 	&	245.50 	&	102-4850	\\
J1018-1642	&	-1.33 	&	0.27 	&	1981.00 	&	422.80 	&	102.5-1420	\\
J1045-4509*	&	-1.24 	&	0.04 	&	4622.00 	&	158.70 	&	147.5-3100	\\
J1048-5832	&	-0.45 	&	0.12 	&	19191.80 	&	1070.70 	&	640-17000	\\
J1110-5637	&	-0.53 	&	0.22 	&	4171.90 	&	684.40 	&	640-3100	\\
J1114-6100	&	0.00 	&	0.87 	&	4053.30 	&	635.10 	&	640-3100	\\
J1116-4122	&	-1.39 	&	0.07 	&	3467.50 	&	220.90 	&	147.5-3100	\\
J1121-5444	&	-1.74 	&	0.10 	&	5562.30 	&	2422.60 	&	151.5-3100	\\
J1136-5525	&	-0.20 	&	0.15 	&	1774.30 	&	178.00 	&	154.24-1382	\\
J1225-6408	&	0.00 	&	0.07 	&	1440.00 	&	44.50 	&	400-1440	\\
J1238+2152	&	0.00 	&	0.16 	&	424.00 	&	20.90 	&	25-400	\\
J1239-6832	&	-1.31 	&	0.15 	&	4164.10 	&	997.20 	&	154.24-3100	\\
J1302-6350	&	0.00 	&	0.17 	&	12342.00 	&	4034.60 	&	640-8356	\\
J1313+0931	&	-0.66 	&	0.16 	&	1507.40 	&	26.30 	&	57-1400	\\
J1320-5359	&	-1.36 	&	0.10 	&	6880.60 	&	2477.00 	&	154.24-3100	\\
J1355-5153	&	-1.54 	&	0.12 	&	1761.20 	&	128.20 	&	147.5-1369	\\
J1440-6344	&	-0.87 	&	0.19 	&	1519.10 	&	46.60 	&	154.24-1400	\\
J1452-6036	&	0.00 	&	0.19 	&	4515.00 	&	294.80 	&	728-3100	\\
J1507-6640	&	-1.23 	&	0.36 	&	2185.90 	&	250.70 	&	400-1460.04	\\
J1512-5759	&	-0.99 	&	0.20 	&	3986.10 	&	327.60 	&	640-3100	\\
J1553-5456	&	-1.46 	&	0.49 	&	2022.30 	&	236.60 	&	400-1459.32	\\
J1645+1012	&	-0.33 	&	0.58 	&	452.10 	&	16.70 	&	59.68-430	\\
J1646-4346	&	-0.81 	&	0.33 	&	5992.80 	&	1059.90 	&	1350-4860	\\
J1652+2651	&	0.00 	&	0.15 	&	920.10 	&	79.70 	&	102.5-800	\\
J1700-3312	&	-0.78 	&	0.18 	&	4160.90 	&	866.80 	&	350-3100	\\
J1700-3611	&	-0.17 	&	0.28 	&	1984.80 	&	198.00 	&	350-1460.04	\\
J1703-1846	&	-0.99 	&	0.15 	&	1720.00 	&	515.00 	&	102.5-1720	\\
J1703-3241	&	-1.34 	&	0.07 	&	11407.00 	&	5731.20 	&	147.5-5124	\\
J1707-4053	&	-1.54 	&	0.31 	&	8063.70 	&	1337.20 	&	147.5-6591	\\
J1709-1640	&	-0.90 	&	0.04 	&	25259.00 	&	2968.40 	&	49.8-22700	\\
J1720-0212	&	-0.79 	&	0.16 	&	1551.20 	&	53.80 	&	102.5-1420	\\
J1720-1633	&	-1.47 	&	0.12 	&	6006.00 	&	587.90 	&	147.5-4850	\\
J1721-3532	&	-0.47 	&	0.06 	&	21938.70 	&	2340.60 	&	1360-17000	\\
J1723-3659	&	0.00 	&	0.23 	&	4016.50 	&	252.30 	&	325-3100	\\
J1730-2304*	&	-1.22 	&	0.09 	&	3334.60 	&	224.90 	&	102-3100	\\
J1733-2228	&	-0.04 	&	0.91 	&	1597.30 	&	136.10 	&	350-1435	\\
J1735-0724	&	-0.08 	&	0.30 	&	1567.50 	&	71.80 	&	102.5-1408	\\
J1736-2457	&	-0.43 	&	0.30 	&	1886.30 	&	105.10 	&	350-1460.04	\\
J1738-2330	&	-0.45 	&	0.89 	&	1794.80 	&	154.70 	&	350-1460.04	\\
J1739-3023	&	0.00 	&	0.03 	&	3487.90 	&	112.90 	&	300-3100	\\
J1740+1000	&	-0.44 	&	0.06 	&	13966.90 	&	1697.30 	&	149-8350	\\
J1741-2733	&	-0.76 	&	0.17 	&	2153.70 	&	188.10 	&	350-1459.32	\\
J1754+5201	&	-0.11 	&	0.22 	&	5052.10 	&	114.90 	&	102.5-4850	\\
J1756-2435	&	-0.35 	&	0.21 	&	3711.50 	&	274.80 	&	606-3100	\\
J1757-2223	&	0.00 	&	0.04 	&	3913.10 	&	301.70 	&	325-3100	\\
J1759-3107	&	-0.94 	&	0.19 	&	4326.00 	&	1017.20 	&	350-3100	\\
J1801-0357	&	-1.15 	&	0.29 	&	1758.80 	&	181.30 	&	102.5-1420	\\
J1801-2451	&	-0.28 	&	0.15 	&	3516.80 	&	149.10 	&	147.5-3100	\\
J1804-2717*	&	-0.27 	&	0.50 	&	1500.60 	&	61.40 	&	350-1400	\\
J1807-2715	&	-1.69 	&	0.37 	&	1959.40 	&	376.00 	&	350-1420	\\
J1808-0813	&	-1.38 	&	0.19 	&	3999.10 	&	546.20 	&	102.5-3100	\\
J1808-2057	&	-0.97 	&	0.19 	&	5905.50 	&	440.50 	&	606-4850	\\
J1810+1744*	&	-0.55 	&	0.29 	&	416.50 	&	13.70 	&	74-400	\\
J1812-1733	&	-0.92 	&	0.18 	&	4256.80 	&	197.80 	&	728-4000	\\
J1816-1729	&	-0.27 	&	0.36 	&	2415.60 	&	514.60 	&	147.5-1606	\\
J1816-2650	&	-0.93 	&	0.28 	&	1952.50 	&	278.70 	&	350-1400	\\
J1817-3618	&	-0.55 	&	0.16 	&	3373.30 	&	242.00 	&	350-3100	\\
J1817-3837	&	-0.54 	&	0.15 	&	4497.70 	&	735.10 	&	350-3100	\\
J1825+0004	&	-1.10 	&	0.30 	&	1729.20 	&	227.60 	&	102.5-1408	\\
J1828-1101	&	-0.14 	&	0.28 	&	5015.80 	&	150.90 	&	1360-4850	\\
J1829-1751	&	-0.95 	&	0.06 	&	11601.90 	&	714.00 	&	147.5-10550	\\
J1834-0426	&	-0.93 	&	0.06 	&	6791.10 	&	572.70 	&	102.5-4850	\\
J1835-0643	&	-1.08 	&	0.07 	&	5227.50 	&	441.00 	&	147.5-4850	\\
J1837-0653	&	-1.08 	&	0.30 	&	4413.70 	&	1196.70 	&	102.5-3100	\\
J1841-0425	&	0.00 	&	0.09 	&	3875.00 	&	182.40 	&	300-3100	\\
J1842-0153	&	-0.37 	&	0.30 	&	3565.80 	&	232.30 	&	728-3100	\\
J1843-0211	&	-0.51 	&	0.29 	&	3903.30 	&	412.70 	&	728-3100	\\
J1845-0434	&	0.00 	&	0.62 	&	2432.00 	&	255.40 	&	606-1606	\\
J1850+0026	&	-0.78 	&	0.22 	&	2330.70 	&	377.10 	&	350-1459.32	\\
J1850+1335	&	-0.53 	&	0.37 	&	1694.20 	&	158.90 	&	102.5-1606	\\
J1852+0031	&	-0.50 	&	0.10 	&	5210.50 	&	80.50 	&	300-4850	\\
J1855-0941	&	-0.54 	&	0.28 	&	1777.80 	&	88.60 	&	350-1460.04	\\
J1901-0906	&	-0.26 	&	0.23 	&	3209.40 	&	74.30 	&	350-3100	\\
J1903+0327*	&	0.00 	&	0.12 	&	5675.10 	&	242.70 	&	74-5000	\\
J1903-0632	&	-1.74 	&	0.07 	&	3500.70 	&	246.70 	&	102.5-3100	\\
J1909+0007	&	-1.08 	&	0.19 	&	1740.40 	&	114.90 	&	102.5-1408	\\
J1910-0309	&	-1.91 	&	0.18 	&	2198.20 	&	574.90 	&	102.5-1420	\\
J1912+2104	&	-0.13 	&	0.25 	&	1721.80 	&	135.70 	&	149-1615	\\
J1921+1948	&	-0.94 	&	0.13 	&	1715.50 	&	131.20 	&	147.5-1420	\\
J1926+0431	&	-1.19 	&	0.21 	&	6965.00 	&	1472.40 	&	102.5-4850	\\
J1946-2913	&	-0.60 	&	0.34 	&	1530.90 	&	90.10 	&	350-1408	\\
J2007+2722	&	0.00 	&	0.08 	&	10216.00 	&	950.40 	&	610-9000	\\
J2013+3845	&	-0.85 	&	0.08 	&	6204.90 	&	394.10 	&	102.5-4850	\\
J2051-0827*	&	-1.37 	&	0.06 	&	26950.00 	&	18220.90 	&	111-2695	\\
J2053-7200	&	-1.16 	&	0.19 	&	3411.80 	&	353.90 	&	151.5-3100	\\
J2116+1414	&	-0.64 	&	0.35 	&	1889.30 	&	282.10 	&	135.25-1420	\\
J2150+5247	&	-0.74 	&	0.13 	&	6646.20 	&	676.10 	&	102.5-4850	\\
J2157+4017	&	-0.46 	&	0.06 	&	5013.80 	&	30.80 	&	100-4850	\\
J2212+2933	&	-0.01 	&	4.28 	&	1591.60 	&	365.30 	&	102.5-1420	\\
J2225+6535	&	-1.14 	&	0.13 	&	2704.20 	&	746.90 	&	59.7-1606	\\
J2325+6316	&	0.00 	&	0.23 	&	1740.60 	&	151.40 	&	102.5-1420	\\

\end{longtable}

\begin{table*}
\centering
\caption{\textbf{Double turn-over spectrum (has a low-frequency turn-over and a high-frequency cut-off), where $\alpha$ is the spectral index for a given frequency range, $\beta$ the smoothness of the turn-over, $\nu_{c}$ and $\nu_{p}$ the cut off frequency and the peak/turn-over frequency. They are followed by corresponding errors. We also mark the MSPs with*. }}
\begin{threeparttable}
\label{tab:a2}
\begin{tabular}{cccccccccc} % four columns, alignment for each
\hline
PSRJ        & $\alpha$& $\alpha$\_err & $\beta$& $\beta$\_err & $\nu_{p}(MHz)$  & $\nu_{p}$\_err(MHz) & $\nu_{c}$(MHz)& $\nu_{c}$\_err(MHz) & Freq.range(MHz) \\
\hline
J0024-7204D*	&	-5.63 	&	0.08 	&	0.21 	&	0.00 	&	689.5	&	40.6	&	4225	&	239.4	&	436-3968	\\
J0030+0451*	&	-2.26 	&	0.02 	&	2.10 	&	1.26 	&	55.1	&	0.2	&	21000	&	16672.1	&	50-2100	\\
J0034-0534*	&	-4.70 	&	0.29 	&	1.30 	&	0.09 	&	84.1	&	0.5	&	16600	&	14507	&	50-1660	\\
J0034-0721	&	-3.44 	&	0.18 	&	0.74 	&	0.05 	&	72.7	&	2	&	14350	&	9435	&	20-1435	\\
J0141+6009	&	-1.47 	&	0.07 	&	2.10 	&	1.64 	&	102	&	9.9	&	6654.4	&	590.7	&	58.6-4850	\\
J0630-2834	&	-1.86 	&	0.03 	&	2.10 	&	0.14 	&	83.5	&	0.4	&	105500	&	57713.6	&	35.1-10550	\\
J0946+0951	&	-5.54 	&	2.19 	&	0.72 	&	0.43 	&	53	&	2.8	&	14080	&	6881.1	&	20-1408	\\
J0953+0755	&	-1.36 	&	0.02 	&	2.10 	&	0.30 	&	53.4	&	2.1	&	26832.9	&	641.5	&	20-22700	\\
J1022+1001*	&	-1.13 	&	0.26 	&	1.36 	&	1.50 	&	73.3	&	16.2	&	8158.3	&	2184.9	&	50-4850	\\
J1239+2453	&	-8.00 	&	5.10 	&	0.41 	&	0.04 	&	71.5	&	2.9	&	22700	&	182092.8	&	20-22700	\\
J1456-6843	&	-2.61 	&	0.10 	&	2.10 	&	0.21 	&	259	&	2.7	&	83560	&	74046.4	&	102-8356	\\
J1532+2745	&	-8.00 	&	6.88 	&	0.40 	&	0.07 	&	69.4	&	8.5	&	1420	&	8199	&	25-1420	\\
J1607-0032	&	-1.93 	&	0.13 	&	1.66 	&	0.24 	&	78.8	&	3.5	&	105500	&	55751.5	&	25-10550	\\
J1741-3016	&	-6.32 	&	0.06 	&	0.28 	&	0.00 	&	947.9	&	26.3	&	3719.2	&	203.8	&	300-3100	\\
J1825-0935	&	-0.92 	&	0.09 	&	2.10 	&	1.27 	&	54.8	&	7.1	&	12061.8	&	986.2	&	25-10550	\\
J1825-1446	&	-3.79 	&	0.08 	&	0.13 	&	0.00 	&	1722.7	&	223.8	&	9265.1	&	969.8	&	325-8000	\\
J1832-1021	&	-5.51 	&	0.05 	&	0.37 	&	0.01 	&	2000	&	1483.6	&	1674.3	&	18.8	&	408-1606	\\
J1840+5640	&	-1.11 	&	0.12 	&	2.10 	&	1.62 	&	44.9	&	2.8	&	6563.7	&	932.3	&	20-4850	\\
J1911-1114*	&	-8.00 	&	7.05 	&	0.21 	&	0.00 	&	139.7	&	8.8	&	7757.3	&	10512.6	&	102-1660	\\
J1932+1059	&	-1.06 	&	0.03 	&	1.63 	&	0.34 	&	63.7	&	6.5	&	50801.4	&	5737.5	&	20-43000	\\
J1955+2908*	&	-3.31 	&	0.94 	&	1.34 	&	0.77 	&	168.8	&	17.6	&	14000	&	9748.2	&	147.5-1400	\\
J2048-1616	&	-1.27 	&	0.04 	&	2.10 	&	2.00 	&	10	&	57.1	&	22700	&	7481.4	&	102.5-22700	\\
J2145-0750*	&	-1.89 	&	0.07 	&	1.61 	&	0.19 	&	108.5	&	5.5	&	51240	&	45404	&	50-5124	\\

\hline
\end{tabular}
\end{threeparttable}
\end{table*}

\begin{table*}
\centering
\caption{\textbf{The pulsars that cannot be accurately fitted by the program. We independently fit each pulsar using the least-squares method, calculate the spectral index $\alpha$ along with its uncertainty $\alpha$\_err, and provide the corresponding frequency range.}}
\begin{threeparttable}
\label{tab:a5}
\begin{tabular}{ccccc} % four columns, alignment for each
\hline
PSRJ       & $\alpha$& $\alpha$\_err & Freq.range(MHz) &  \\
\hline
J1115+5030	&	-0.86 	&	0.24 	&	20-1435	&	\\
J1543-0620	&	-1.11 	&	0.31 	&	25-1435	&	\\
J1921+2153	&	-1.52 	&	0.30 	&	20-1420	&	\\
J1948+3540	&	-1.48 	&	0.29 	&	102.5-8600	&	\\
J1955+5059	&	-0.93 	&	0.23 	&	59.66-4920	&	\\
J0014+4746	&	-0.78 	&	0.27 	&	59.14-1435	&	\\
J0214+5222	&	-1.38 	&	0.62 	&	50-820	&	\\
J0332+5434	&	-1.29 	&	0.11 	&	25-43000	&	\\
J0459-0210	&	-2.08 	&	0.22 	&	102.5-1360	&	\\
J0534+2200	&	-1.36 	&	0.81 	&	102.5-1435	&	\\
J0659+1414	&	-0.84 	&	0.19 	&	85-8600	&	\\
J0922+0638	&	-1.40 	&	0.12 	&	20-10550	&	\\
J1136+1551	&	-1.34 	&	0.10 	&	20-22700	&	\\
J1645-0317	&	-1.61 	&	0.11 	&	49.8-22700	&	\\
J1741-0840	&	-1.66 	&	0.51 	&	102.5-4850	&	\\
J1848-0123	&	-1.50 	&	0.20 	&	102.5-10550	&	\\
J1900-2600	&	-1.61 	&	0.25 	&	147.5-4820	&	\\
J1903+0135	&	-1.46 	&	0.51 	&	350-4850	&	\\
J1922+2110	&	-1.59 	&	0.66 	&	135.25-1435	&	\\
J2022+5154	&	-0.81 	&	0.11 	&	65-43000	&	\\
J2108+4441	&	-1.03 	&	0.45 	&	135.25-1435	&	\\
J2113+4644	&	-1.44 	&	0.31 	&	102.5-10550	&	\\
J2354+6155	&	-0.98 	&	0.27 	&	102.5-10550	&	\\
J0454+5543	&	-0.99 	&	0.19 	&	25-10550	&	\\
J0139+5814	&	-1.62 	&	0.23 	&	102.5-10550	&	\\
J0152-1637	&	-1.51 	&	0.18 	&	35.1-1440	&	\\
J0157+6212	&	-0.82 	&	0.56 	&	102.5-1408	&	\\
J0528+2200	&	-1.25 	&	0.15 	&	39-22700	&	\\
J1509+5531	&	-1.30 	&	0.26 	&	20-4820	&	\\
J1543+0929	&	-1.87 	&	0.27 	&	53-4850	&	\\
J1740+1311	&	-1.32 	&	0.23 	&	59.65-4850	&	\\
J1752-2806	&	-1.83 	&	0.19 	&	61-22700	&	\\
J1759-2205	&	-1.65 	&	0.43 	&	147.5-3100	&	\\
J1824-1945	&	-1.36 	&	0.29 	&	147.5-3100	&	\\
J1847-0402	&	-1.62 	&	0.41 	&	102.5-3100	&	\\
J1901+0331	&	-2.04 	&	0.52 	&	102.5-3100	&	\\
J1902+0556	&	-1.39 	&	0.52 	&	102.5-3100	&	\\
J1916+1312	&	-1.70 	&	0.38 	&	102.5-4850	&	\\
J1917+1353	&	-1.67 	&	0.33 	&	102.5-4850	&	\\
J1919+0021	&	-1.95 	&	0.88 	&	135.25-1410	&	\\
J1954+2923	&	-0.47 	&	0.30 	&	25-4920	&	\\
J2018+2839	&	-1.27 	&	0.15 	&	25-22700	&	\\
J2257+5909	&	-1.33 	&	0.46 	&	102.5-4820	&	\\
J2326+6113	&	-0.91 	&	0.52 	&	102.5-1420	&	\\

\hline
\end{tabular}
\end{threeparttable}
\end{table*}

\newpage
\bibliography{references}{}
\bibliographystyle{aasjournal}

%%%%%%%%%%%%%%%%% APPENDICES %%%%%%%%%%%%%%%%%%%%%

\end{document}